\documentclass[twocolumn]{aastex61}
\pdfoutput=1 
\usepackage{amsmath,amstext}
\usepackage[T1]{fontenc}
\usepackage{apjfonts} 
\usepackage[figure,figure*]{hypcap}
\usepackage{amsfonts}
\usepackage{amssymb}
\usepackage{tabularx}
\usepackage{xspace}

\newcommand{\lya}{Ly-$\alpha$\xspace}

\shorttitle{First SHIZUCA results. II. Spectroscopy}
\shortauthors{Ch. Curtin et al.}

\begin{document}

\title{First release of high redshift superluminous supernovae from \\
the Subaru high-z supernova campaign (SHIZUCA).\\
II. Spectroscopic properties.}

\author{Chris Curtin}
\affiliation{Centre for Astrophysics \& Supercomputing, Swinburne University of Technology, Hawthorn, VIC 3122, Australia}
\affiliation{ARC Centre of Excellence for All-Sky Astrophysics (CAASTRO)}
\email{ccurtin@swin.edu.au}

\author{Jeff Cooke}
\affiliation{Centre for Astrophysics \& Supercomputing, Swinburne University of Technology, Hawthorn, VIC 3122, Australia}
\affiliation{ARC Centre of Excellence for All-Sky Astrophysics (CAASTRO)}
\author{Takashi J. Moriya}
\affiliation{National Astronomical Observatory of Japan, National Institutes of Natural Sciences, 2-21-1 Osawa, Mitaka, Tokyo 181-8588, Japan}

\author{Masayuki Tanaka}
\affiliation{National Astronomical Observatory of Japan, National Institutes of Natural Sciences, 2-21-1 Osawa, Mitaka, Tokyo 181-8588, Japan}

\author{Robert M. Quimby}
\affiliation{Department of Astronomy / Mount Laguna Observatory, San Diego State University, 5500 Campanile Drive, San Diego, CA, 92812-1221, USA}
\affiliation{Kavli Institute for the Physics and Mathematics of the Universe (WPI), The University of Tokyo Institutes for Advanced Study, The University of Tokyo, 5-1-5 Kashiwanoha, Kashiwa, Chiba 277-8583, Japan}

\author{Stephanie R. Bernard}
\affiliation{School of Physics, University of Melbourne, Parkville VIC 3010, Australia}
\affiliation{ARC Centre of Excellence for All-Sky Astrophysics (CAASTRO)}

\author{Llu\'is Galbany}
\affiliation{PITT PACC, Department of Physics and Astronomy, University of Pittsburgh, Pittsburgh, PA 15260, USA}

\author{Ji-an Jiang}
\affiliation{Institute of Astronomy, Graduate School of Science, The University of Tokyo, 2-21-1 Osawa, Mitaka, Tokyo 181-0015, Japan}

\author{Chien-Hsiu Lee}
\affiliation{Subaru Telescope, NAOJ, 650 N Aohoku Pl., Hilo, HI 96720, USA}

\author{Keiichi Maeda}
\affiliation{Department of Astronomy, Kyoto University, Kitashirakawa-Oiwake-cho, Sakyo-ku, Kyoto 606-8502, Japan}
\affiliation{Kavli Institute for the Physics and Mathematics of the Universe (WPI), The University of Tokyo Institutes for Advanced Study, The University of Tokyo, 5-1-5 Kashiwanoha, Kashiwa, Chiba 277-8583, Japan}

\author{Tomoki Morokuma}
\affiliation{Institute of Astronomy, Graduate School of Science, The University of Tokyo, 2-21-1 Osawa, Mitaka, Tokyo 181-0015, Japan}

\author{Ken'ichi Nomoto}
\affiliation{Kavli Institute for the Physics and Mathematics of the Universe (WPI), The University of Tokyo Institutes for Advanced Study, The University of Tokyo, 5-1-5 Kashiwanoha, Kashiwa, Chiba 277-8583, Japan}

\author{Giuliano Pignata}
\affiliation{Departamento de Ciencias F\'isicas, Universidad Andres Bello, Avda. Rep\'ublica 252, Santiago, 8320000, Chile}
\affiliation{Millennium Institute of Astrophysics (MAS), Nuncio Monse\~nor S\'otero Sanz 100, Providencia, Santiago, Chile}

\author{Tyler Pritchard}
\affiliation{Centre for Astrophysics \& Supercomputing, Swinburne University of Technology, Hawthorn, VIC 3122, Australia}

\author{Nao Suzuki}
\affiliation{Kavli Institute for the Physics and Mathematics of the Universe (WPI), The University of Tokyo Institutes for Advanced Study, The University of Tokyo, 5-1-5 Kashiwanoha, Kashiwa, Chiba 277-8583, Japan}

\author{Ichiro Takahashi}
\affiliation{Kavli Institute for the Physics and Mathematics of the Universe (WPI), The University of Tokyo Institutes for Advanced Study, The University of Tokyo, 5-1-5 Kashiwanoha, Kashiwa, Chiba 277-8583, Japan}

\author{Masaomi Tanaka}
\affiliation{Astronomical Institute, Tohoku University, 6-3 Aramaki Aza-Aoba, Aoba-ku, Sendai 980-8578, Japan}

\author{Nozomu Tominaga}
\affiliation{Department of Physics, Faculty of Science and Engineering, Konan University, 8-9-1 Okamoto, Kobe, Hyogo 658-8501, Japan}
\affiliation{Kavli Institute for the Physics and Mathematics of the Universe (WPI), The University of Tokyo Institutes for Advanced Study, The University of Tokyo, 5-1-5 Kashiwanoha, Kashiwa, Chiba 277-8583, Japan}

\author{Masaki Yamaguchi}
\affiliation{Institute of Astronomy, Graduate School of Science, The University of Tokyo, 2-21-1 Osawa, Mitaka, Tokyo 181-0015, Japan}

\author{Naoki Yasuda}
\affiliation{Kavli Institute for the Physics and Mathematics of the Universe (WPI), The University of Tokyo Institutes for Advanced Study, The University of Tokyo, 5-1-5 Kashiwanoha, Kashiwa, Chiba 277-8583, Japan}


\begin{abstract}  

We present Keck spectroscopic observations of three probable high redshift superluminous supernovae (SLSNe) from the Subaru HIgh-Z sUpernova CAmpaign (SHIZUCA), confirming redshifts of 1.851, 1.965 and 2.399.  The host galaxies were selected for transient monitoring from multi-band photometric redshifts.  The supernovae are detected during their rise, and the classically scheduled spectra are collected near maximum light.  The restframe far-ultraviolet (FUV; $\sim$1000\text{\AA}--2500\text{\AA}) spectra include a significant host galaxy flux contribution and we compare our host galaxy subtracted spectra to UV-luminous SNe from the literature.  While the signal-to-noise ratios of the spectra presented here are sufficient for redshift confirmation, supernova spectroscopic type confirmation remains inconclusive.  The success of the first SHIZUCA Keck spectroscopic follow-up program demonstrates that campaigns such as SHIZUCA are capable of identifying high redshift SLSNe with sufficient accuracy, speed and depth for rapid, well-cadenced and informative follow-up. 

\end{abstract}


\section{INTRODUCTION}\label{intro}

It is now known that some supernovae exceed an absolute magnitude of M $\simeq-$21, giving rise to a luminosity class of supernovae termed superluminous supernovae (SLSNe; \citealp{quimby11,galyam12,nicholl15Ics,moriya2018slsne}).  SLSNe are also more luminous in the far-ultraviolet (FUV;  \mbox{$\sim$1000\text{\AA}--2500\text{\AA}}) relative to other supernovae \citep{quimby11,cooke12hiz,howell13hizslsne,yan1716apd,Yan17egm}.  SLSNe are exceedingly rare, occurring at $\sim0.001\times$ the frequency of general core-collapse supernovae \citep{quimby13rates,prajs16}.  But these transients promise to be powerful probes of the early universe as they are already visible in ground based observations to redshifts of $z=4$ and greater \citep{cooke12hiz,mould17,Takashi}.

Recent improvements to survey astronomy, such as large-format CCD mosaics, are enabling more wide-area surveys to operate, such as the Hyper Suprime-Cam Subaru Strategic Program (HSC-SSP; \citealp{hscssp}), the Dark Energy Survey (DES; \citealp{des}), the Zwicky Transient Facility \citep{ztf17} and Pan-STARRS \citep{kaiser10panstarrs}.  These surveys monitor the vast amounts of sky required to detect SLSNe with a reasonable frequency.  In addition these surveys are broad enough to allow boutique surveys such as the Subaru HIgh-Z sUpernova CAmpaign (SHIZUCA; \citealp{Takashi}) and the Survey Using DECam for Superluminous Supernovae (SUDSS; \citealp{smith16sudss}) to exercise more focused monitoring techniques and detect SLSNe with higher efficiency.

There are advantages to surveying for SLSNe at high redshift ($z\gtrsim 2$).  Massive progenitors are invoked in all the explosion mechanisms used to explain SLSNe \citep{galyam12,moriya2018slsne}, suggesting that the rate of SLSNe per unit volume varies with redshift as the cosmic star formation rate, rising and perhaps peaking at $z\sim2$ \citep{CSFR}.  This is consistent with rates measured up to this redshift \citep{quimby13rates,prajs16}.  Some if not all types of SLSNe show a preference for low-metallicity hosts \citep{lunnan14,leloudas15,angus16,schulze18}, suggesting that SLSNe may require low-metallicity progenitors and that the volumetric rate of SLSNe may continue to increase with redshift beyond the peak of star formation into epochs of universally low metallicity.  The measured rate at $z\sim$ \mbox{2--4} is consistent with a continuing increase beyond $z\sim2$ \citep{cooke12hiz}, though this measurement is based on two events.  

Another advantage of looking to high redshift is the ability to sample the restframe-FUV of SLSNe with optical facilities.  FUV spectra of low redshift supernovae are rare because they can only be collected from space-based telescopes.  Studies of such spectra have been done \citep{fransson02,fransson05,fransson14,panagia07uvspectra,bufano09swiftspectra}, including SLSNe \citep{yan1716apd,Yan17egm,quimby18slsnIspec}.  Due in part to their excessive UV luminosity and to model-predicted UV indicators of certain explosion mechanisms \citep{mazzali16}, there are programs focused on collecting FUV spectra of SLSNe in particular \citep{quimby14hstpropslsnII,quimby16hstpropslsn}.  Still, given the low rate of SLSNe, few such spectra have been collected to date.  At high redshift this critical UV information is pushed into the optical bands, enabling this wavelength region to be explored with ground-based observations.  A number of SLSN FUV analyses of have been accomplished in this way \citep{berger1211bam,howell13hizslsne,pan15e2mlf,smith18c2nm}.

High \'entendu surveys like SHIZUCA are already capable of identifying supernovae to $z>4$ \citep{Takashi,mould17}.  Spectroscopic follow-up of high redshift supernovae must be carried out on 8-meter class telescopes due to their intrinsically faint apparent magnitudes (often m$_{r}\gtrsim24$ for $z\gtrsim2$).

Here we present spectra obtained with Keck of three probable high redshift SLSNe.  This paper summarizes the first spectroscopic follow-up program of SHIZUCA.  It has been written in parallel with \citet{Takashi}, the first photometric analysis of the SHIZUCA program, hereafter M19.  We summarize the observations in Section~2.  In Section~3 we present the spectra and their redshift measurements.  In Section~4 we discuss our analysis of the observations and we present our conclusions in Section~5.  All calculations in this paper assume a $\Lambda$CDM cosmology with $H_{0}=70$km~s$^{-1}$~Mpc$^{-1}$, $\Omega_{M}=0.3$ and $\Omega_{\Lambda}=0.7$.  All magnitudes are AB and all wavelengths are quoted in restframe \text{\AA}ngstr\"oms unless otherwise specified.


\section{OBSERVATIONS}\label{obs}

The COSMOS field \citep{cosmos} of the HSC-SSP\footnote{General information for the HSC-SSP, such as location, cadence, and data products can be found at: http://hsc.mtk.nao.ac.jp/ssp/ and the associated links.} supplies the photometry for this work and consists of a 1.8deg$^{2}$ field-of-view Hyper Suprime-Cam pointing imaged in five filters ($grizy$).  The first SHIZUCA season was active on the COSMOS field from 2016 November to 2017 May.

SHIZUCA uses COSMOS2015 photometric redshifts \citep{laigle16} and \texttt{MIZUKI} redshifts \citep{tanaka15} to identify potential hosts of high redshift transients from HSC-SSP photometry.  The per-epoch depth of observation from the HSC-SSP using the 8.2m Subaru telescope (m$_{i}\lesssim 26.5$) enables SHIZUCA to monitor the fluxes of high redshift sources in individual observing epochs.  Any flux variations observed in these sources are then analyzed in the context of the photometric redshift of the host.  Flux variations are identified as SLSN candidates using criteria such as non-recurrence, duration, light curve shape, peak magnitude and color evolution.  Details of the SHIZUCA photometric analysis and relevant \texttt{MIZUKI} redshift probability distributions are presented and discussed in M19.  

Follow-up spectroscopy of select SHIZUCA photometric high redshift SLSN candidates was acquired on 2016 December 28 and 2017 March \mbox{22--23} using the Low Resolution Imaging Spectrometer \citep[LRIS;][]{oke95,steidel04} on the \mbox{Keck-I} telescope.  These data were obtained using the 400/3400 grism on the blue arm and the 400/8500 grating on the red arm separated at $\sim$5600\text{\AA} using the D560 dichroic.  The CCDs were read out using 2$\times$2 binning with a spectral resolution of $\sim$500km~s$^{-1}$.  The full width at half-maximum (FWHM) seeing ranged from \mbox{$0\farcs9$--$1\farcs1$} and normal atmospheric extinction increased from light cirrus.  During the March observations the blue shutter of LRIS failed and was fixed to remain open, and the trapdoor was used in its place.  The exposures were set at 1200s on the blue side and 1179s on the red side with the difference chosen to allow the CCDs to finish reading out at the same time.

For follow-up we targeted SHIZUCA photometric candidates near maximum light projected to be m$_{r}\lesssim 25.5$ during observation.  This magnitude limit aims to achieve combined transient and host continuum signal-to-noise ratios (S/N) near restframe 1500\text{\AA} of S/N $\sim$ \mbox{3--5}pix$^{-1}$ for integrations of $\sim$2 hours.  SLSNe usually arise in star-forming galaxies such as Lyman break galaxies (LBGs), and these S/N are sufficient to reliably measure redshifts via identification of strong UV absorption features from the interstellar medium along with Lyman-$\alpha$ (Ly-$\alpha$) emission when present and the Ly-$\alpha$ forest when visible \citep{Steidel98}.  However the low S/N combined with systematics in the internal flats increase flux uncertainties at wavelengths shortward of 3600\text{\AA}, observer-frame.

On 2016 December 28, 3 SHIZUCA photometric candidates were eligible for follow-up of which 2 were targeted in the available time, HSC16adga and HSC16aaqc.  On 2017 March 22--23 another 3 SHIZUCA photometric candidates made up the primary target list: HSC17auzg, HSC17dbpf and HSC17cywe.  We followed-up each of these targets with sufficient time remaining to follow-up one backup target, HSC17davs, below our specified magnitude limit.  A spectrum of such low S/N can still yield a measurable redshift and even supernova type information if it includes identifiable strong emission features.

The SHIZUCA photometric candidates are transient phenomena detected photometrically in or near host galaxies with high photometric redshifts (see M19).  There are cases of photometric redshift confusion in which separate, discrete photometric redshift probabilities persist for a single galaxy.  For example, the decrement in the restframe UV continuum of a $z\sim3$ LBG caused by the Ly-$\alpha$ forest can be photometrically mimicked by the Balmer-break in the restframe optical continuum of a $z\sim0.3$ galaxy \citep{steidel03,cooke06lbgs}.
Because the number of targets available for follow-up in a narrow observation window is limited, candidates observed in hosts exhibiting photometric redshift confusion with a reasonable probability of being high redshift are not disqualified from our target lists.  However to minimize the loss of time observing low redshift transients, we inspect the raw 2-D spectra of each target as they are read out for any obvious emission features.  Often one or two emission features are sufficient to differentiate between high and low redshift probabilities.  If a target is determined to be low redshift during observation, the target is aborted and the next target is acquired.

From the 6 SHIZUCA photometric candidates followed up, 3 arise in $z>1.8$ hosts which can be reliably matched with a LBG reference spectrum (see Table~\ref{sources} and Section~\ref{sec:redux}). Two candidates, HSC16aaqc and HSC17cywe, were found to be low redshift through the positive association of emission features from the raw 2-D spectra to commonly observed transitions in low redshift galaxies (e.g., [O\textsc{ii}] $\lambda\lambda$3727, H-$\beta$ and [O\textsc{iii}] $\lambda$5007).  The spectrum of one target, HSC17davs and host, recorded no detectable signal or emissions.  None of the spectra exhibited obvious signs of being other types of high redshift transients, namely active galactic nuclei (AGN) or tidal disruption events (TDEs).

Among the spectra we use as templates for comparison are those of the SLSN-II, LSQ15abl \citep{lsq15abl}. The formal presentation and analysis of these spectra is part of a work in progress.  They were collected with the Hubble Space Telescope (HST) using the Cosmic Origins Spectrograph (COS; \citealp{cos}) and the Space Telescope Imaging Spectrograph (STIS; \citealp{stis}).  The reduced spectra are from the Mikulski Archive for Space Telescopes (MAST) and have been adjusted for scale, noise and redshift. 

\begin{table}
\caption{SHIZUCA Follow-up Targets}
\centering
\begin{tabular}{ccccccccc}
\hline
\hline
SN-ID & Exp Time & Photo-$z^a$ & Spec-$z$ & $\lambda$-Range$^b$ \\
 & blue ; red (s) &  &&(\text{\AA})\\
\hline
2016 DEC & & &
\\
HSC16adga & 7200;7074 & $2.26^{+0.25}_{-0.30}$ & 2.399 & 942--2765 \\
HSC16aaqc & 2400;2358 & 1.25$^{+1.20}_{-0.03}$ & $\sim$0.36$^c$ & --
\\
\hline
2017 MAR & & &
\\
HSC17auzg & 8400;8253 & $1.65^{+0.06}_{-0.08}$ & 1.965 & 1080--3170 \\
HSC17dbpf & 6000;5895 & $2.25^{+0.08}_{-0.53}$ & 1.851 & 1123--3297
\\
HSC17cywe & 2400;2358 & 0.45,3.26$^d$ & $\sim$0.46$^c$ & -- 
\\
HSC17davs & 4800;4716 & 3.34$^{+1.27}_{-2.75}$ & --$^e$ & -- \\
\hline
\end{tabular}
\\
    \small{$^a$ COSMOS2015 unless otherwise specified\\
    $^b$ Restframe, derived from Spec-$z$\\
	$^c$Derived using estimated wavelengths of emission features in raw 2-D spectra\\
	$^d$Two alternative \texttt{MIZUKI} photo-z peaks.  COSMOS2015 photo-z not available\\
	$^e$No signal detected in the reduced imagery}

\label{sources}
\end{table}


\section{DATA REDUCTION AND ANALYSIS}\label{sec:redux}

%
%
\subsection{HSC16adga}

\begin{figure}
\includegraphics[height=6.7cm]{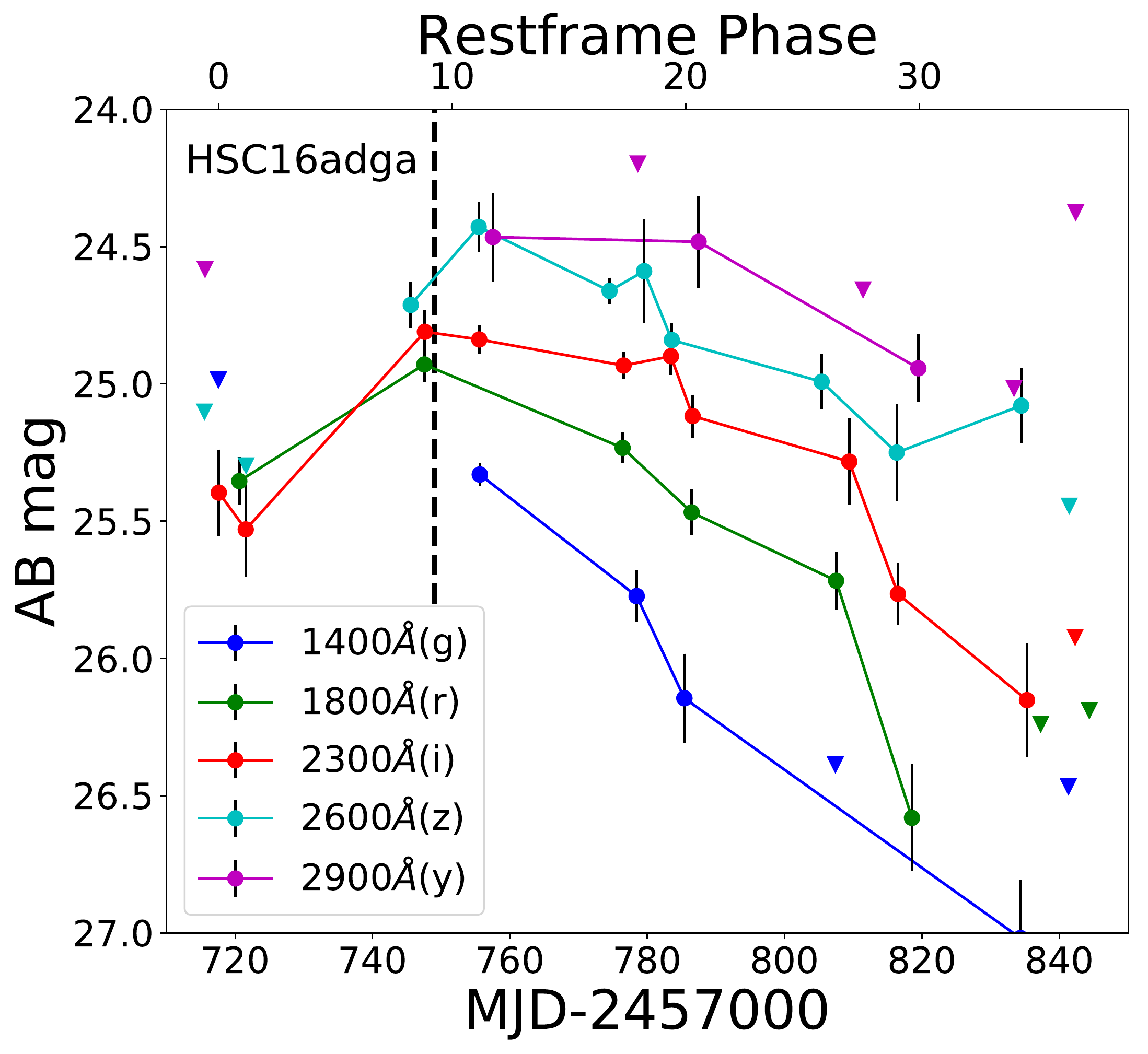}
    \includegraphics[height=6.7cm]{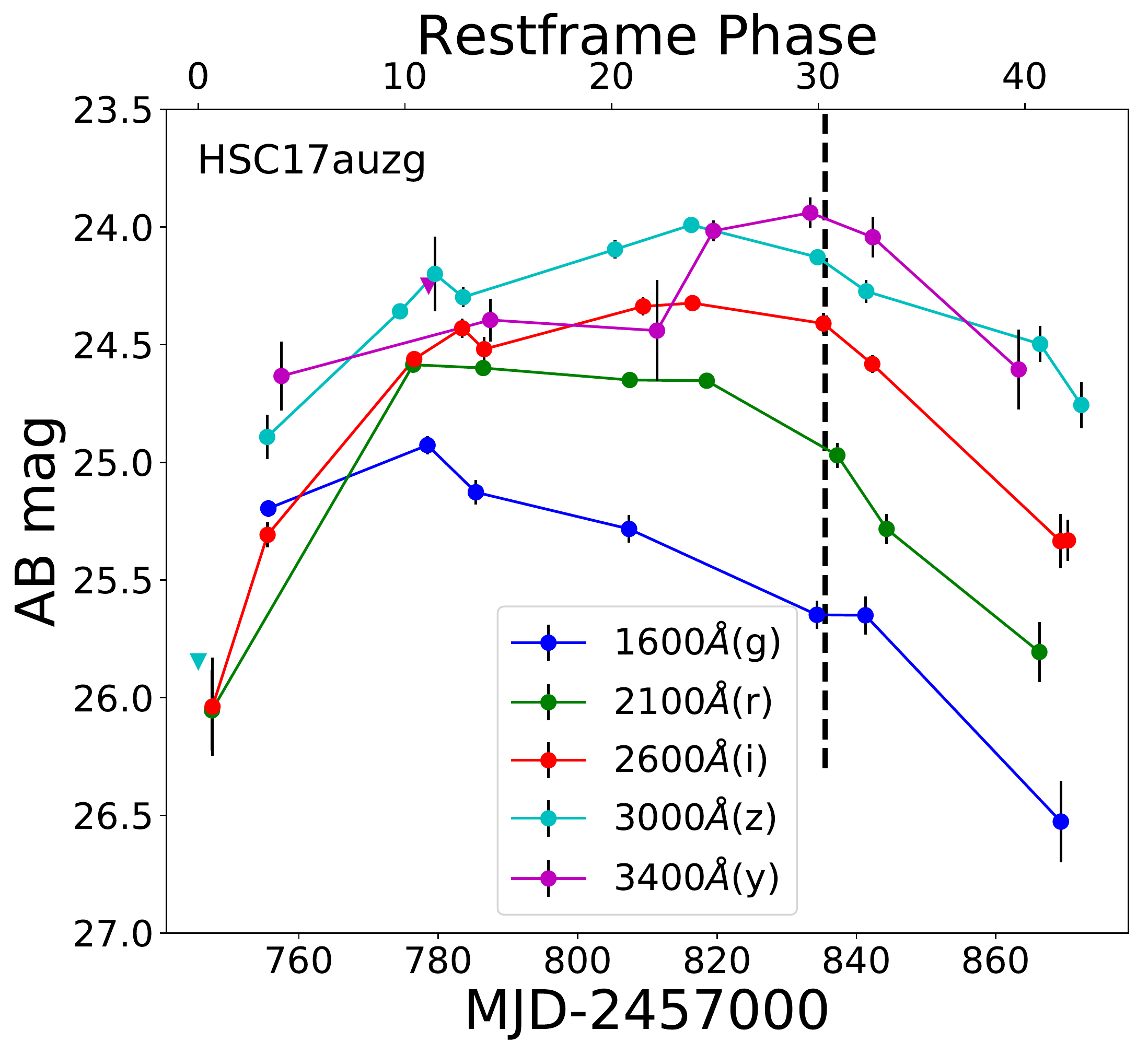}
    \includegraphics[height=6.7cm]{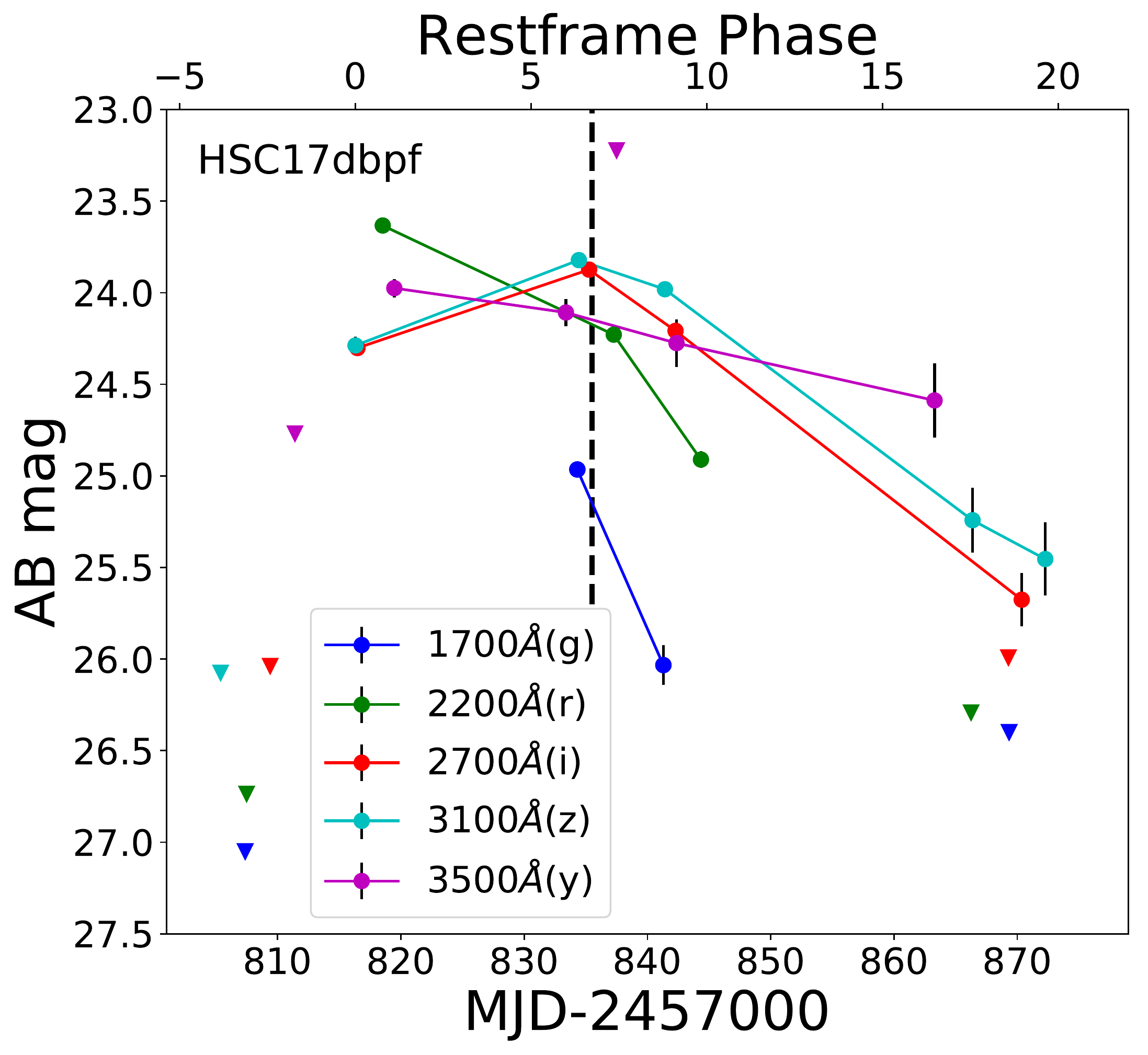}
   \caption{The HSC-$grizy$ light curves of HSC16adga (top), HSC17auzg (middle) and HSC17dbpf (bottom).  Triangles indicate upper limits and errors are 1$\sigma$.  The restframe timescales are relative to the dates of detection and the effective restframe wavelengths of the filters are given in the legends using $z=2.399,$ 1.965 and 1.851 respectively.  Dates that spectra were acquired are identified by dashed black lines.  K-corrected absolute magnitudes are discussed in M19.}
    \label{fig:16adga_phot}
\end{figure}

HSC16adga (SN-2016jhm) was first detected by SHIZUCA on MJD~57715 at coordinates (RA, Dec) = (10:02:20.12, +02:48:43.4).  The transient is detected in a star-forming host with a COSMOS photo-$z$ of $2.26^{+0.25}_{-0.30}$ and a \texttt{MIZUKI} photo-$z$ peaking at $2.19$.  It is observed offset by 0\farcs36$\pm$0\farcs104 ($\sim$3kpc) from the host galaxy flux centroid, helping to rule out an AGN or a TDE.  The light curve of HSC16adga is shown in Figure~\ref{fig:16adga_phot}.

\begin{figure}
\begin{center}
\includegraphics[height=6.7cm]{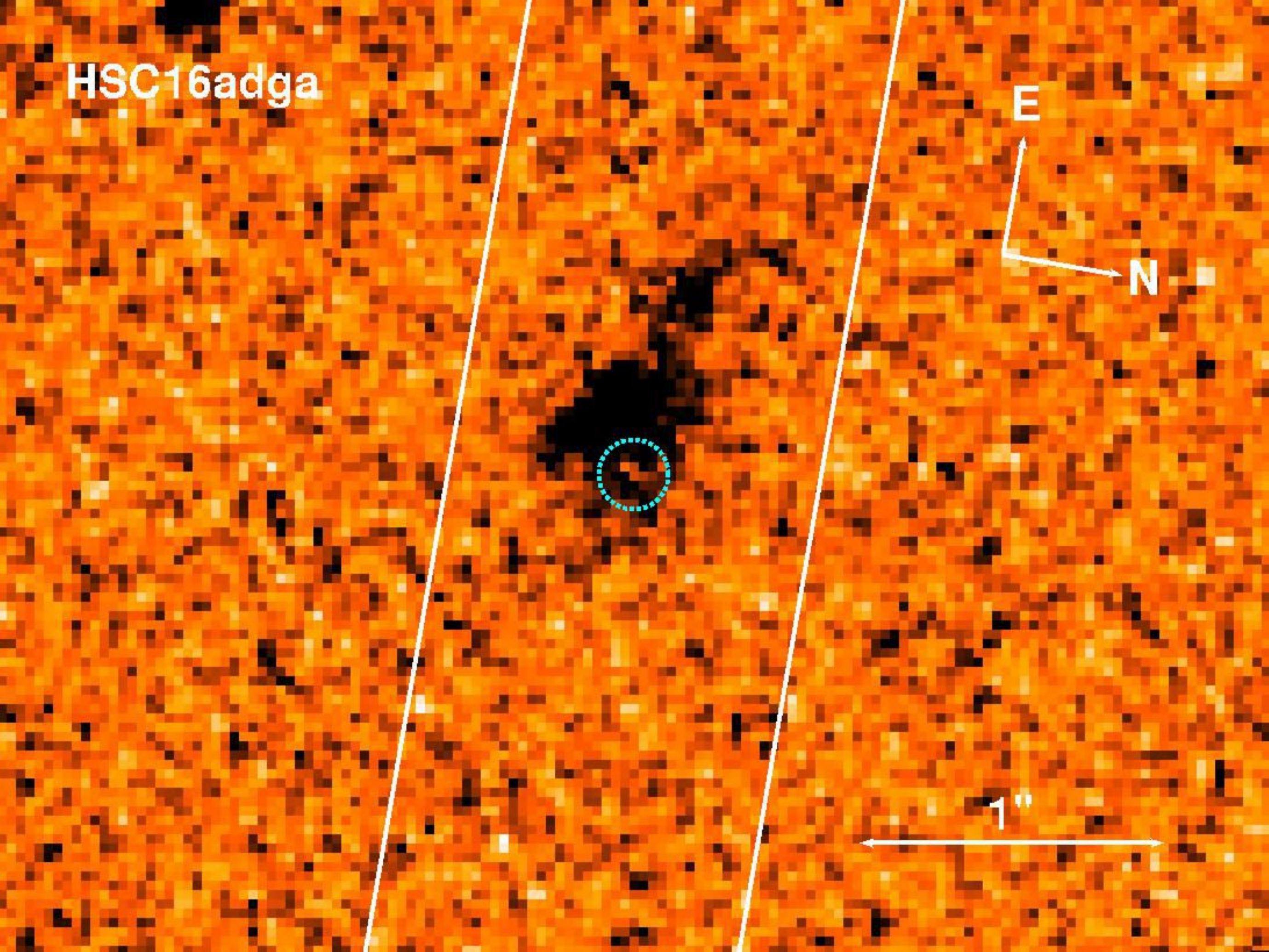}
\includegraphics[height=6.7cm]{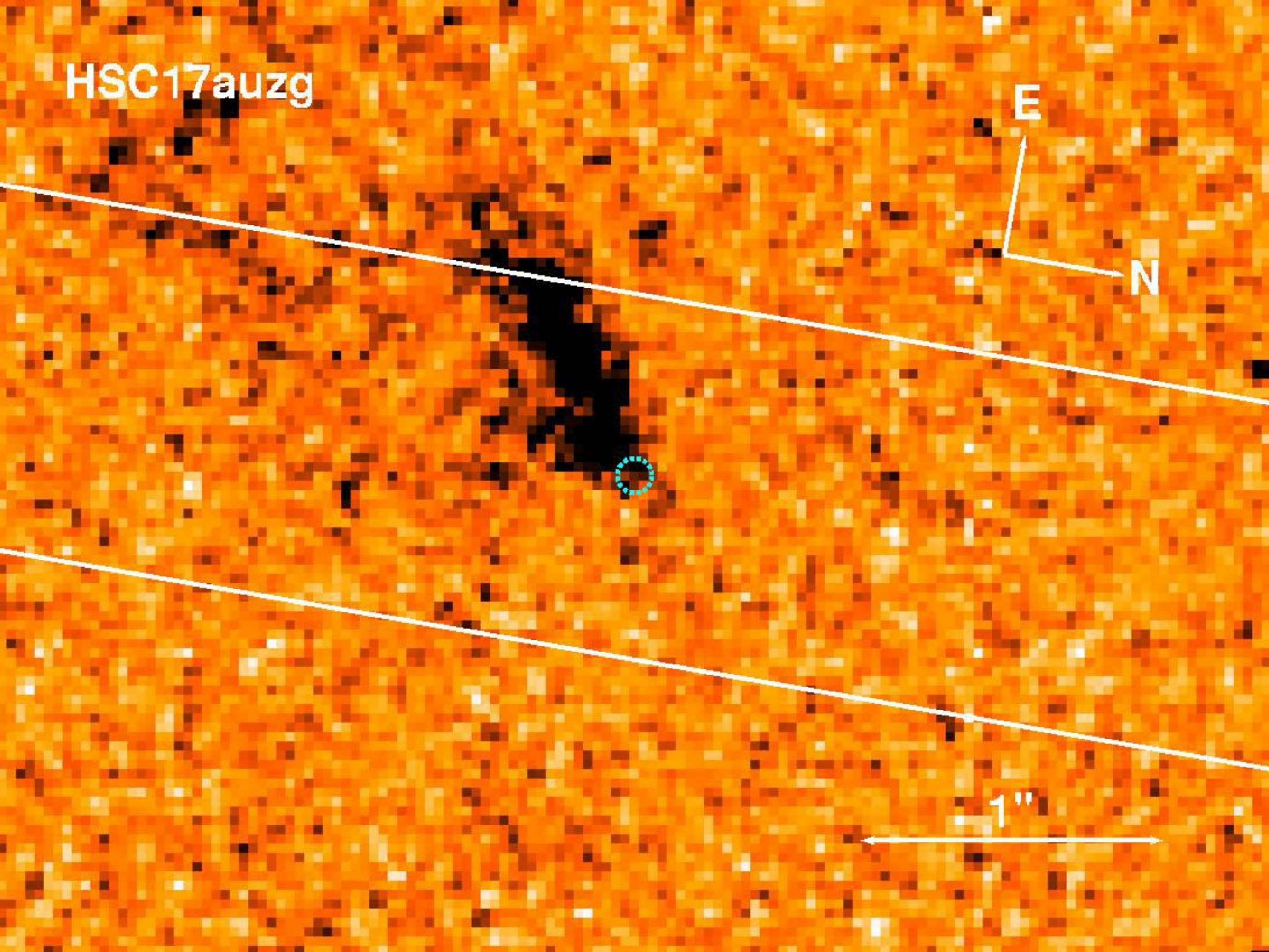}
\includegraphics[height=6.7cm]{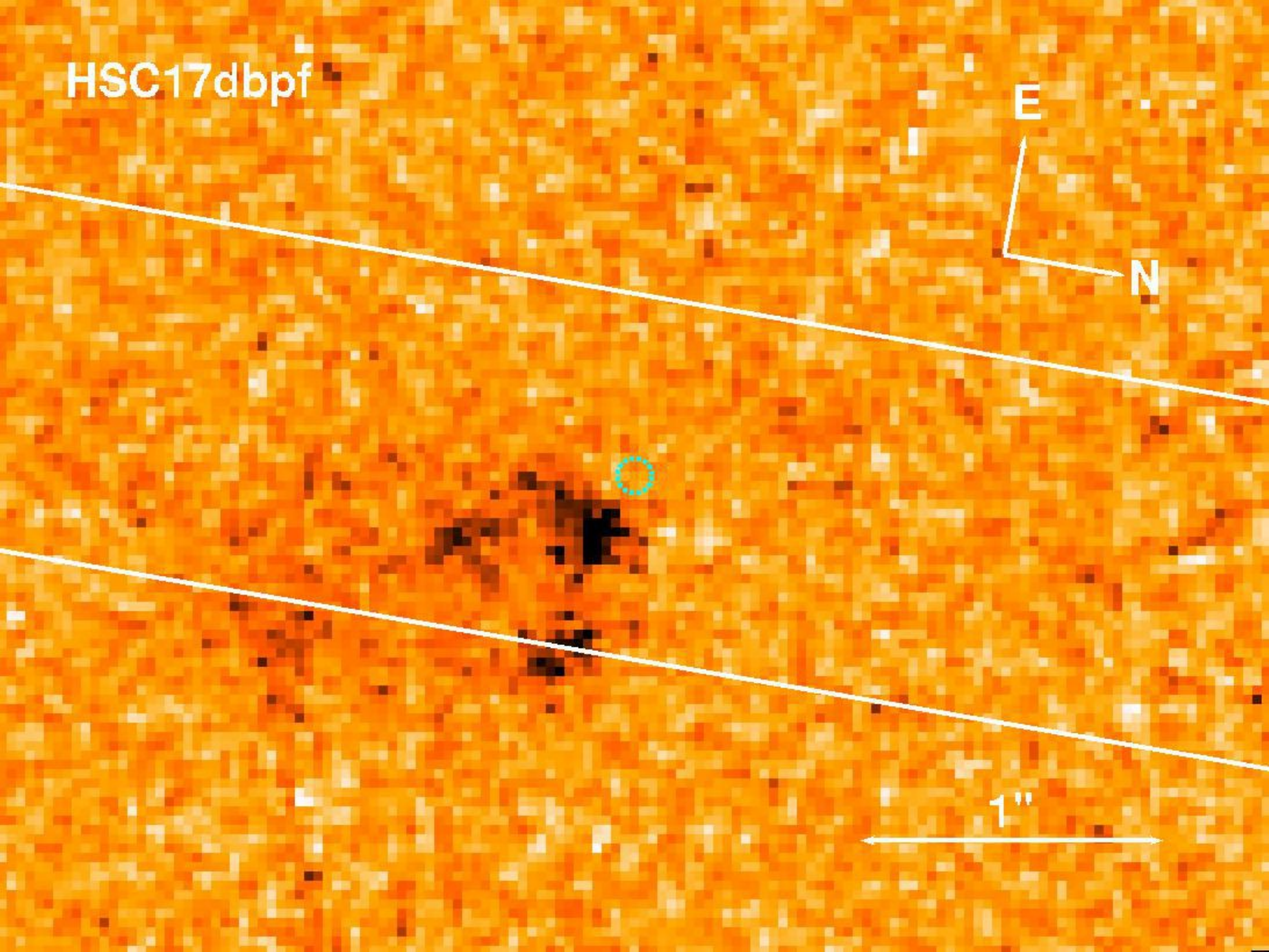}
\end{center}
\caption{Maps of the spectral slits used for the observations of HSC16adga (top), HSC17auzg (middle) and HSC17dbpf (bottom).  The slits are projected onto publicly available HST images of the hosts in quiescence.  The open circles mark the 1$\sigma$ regions of the transient locations as reported by SHIZUCA (see M19).}
\label{fig:hst}
\end{figure}

\begin{figure*}
	\includegraphics[width=\textwidth]{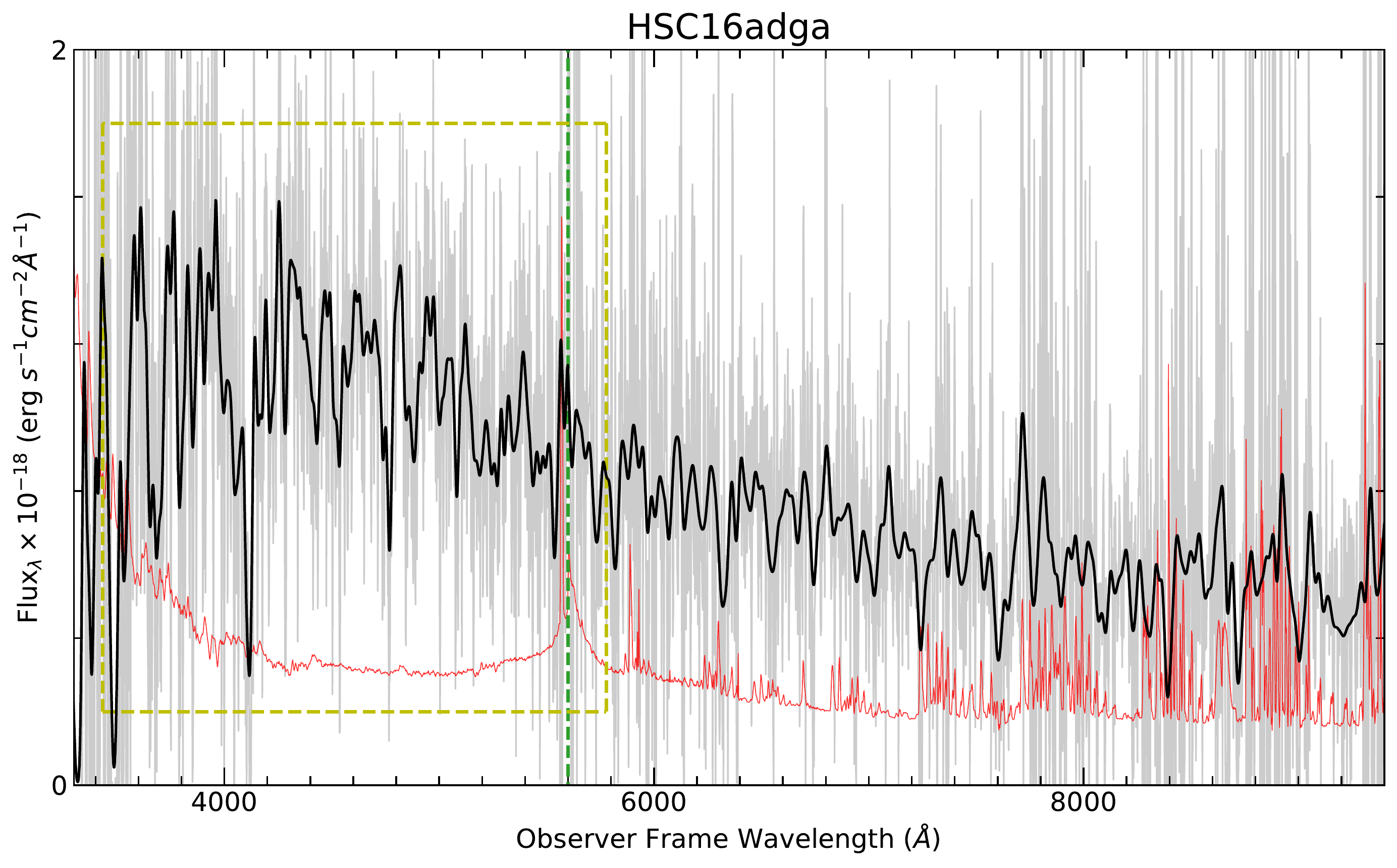}
    \includegraphics[width=\textwidth]{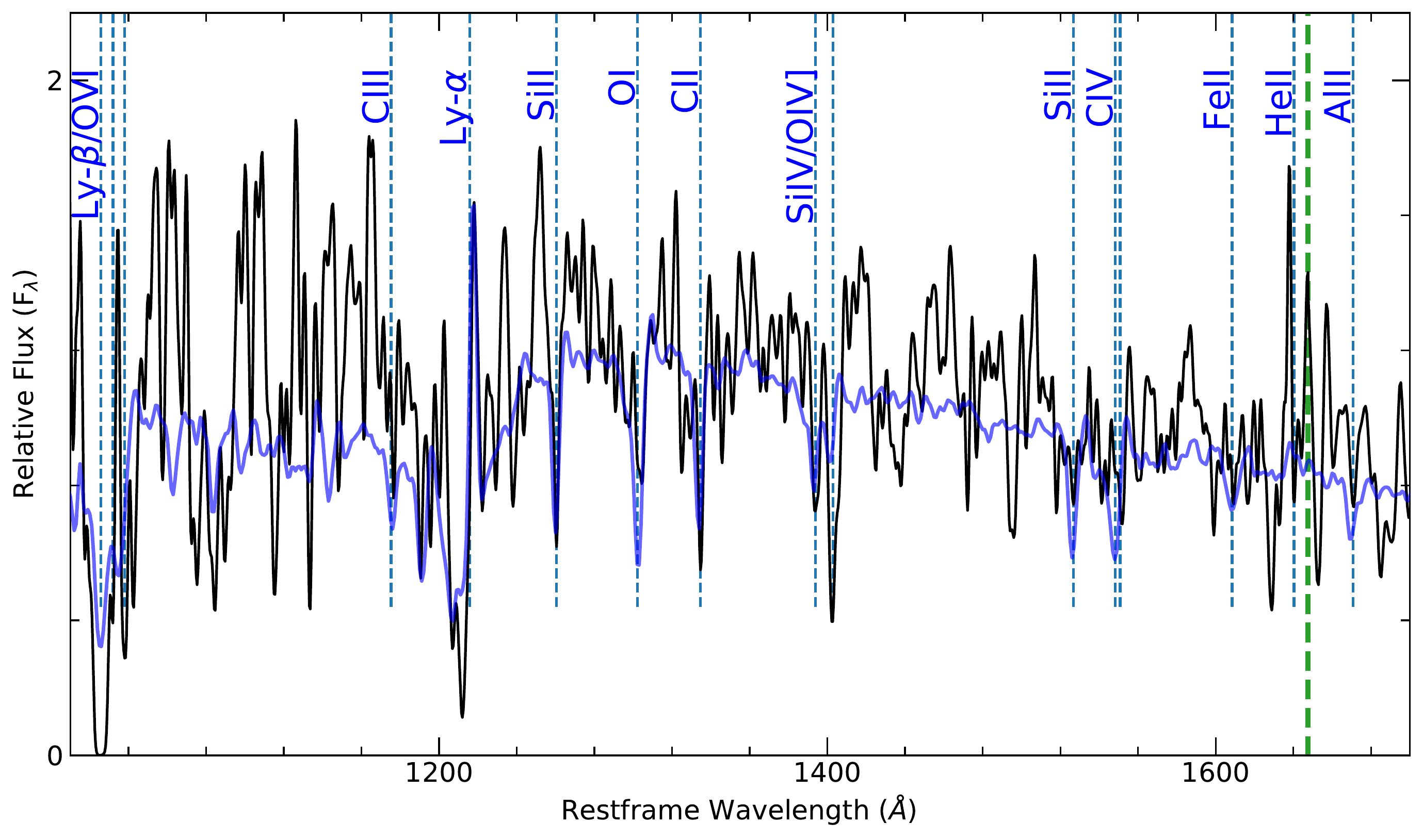}
   \caption{\textit{Top:} Flux-calibrated observer-frame 1-D spectrum of HSC16adga, shown in gray.  A smoothed spectrum is overlaid in black.  The 1$\sigma$~pix$^{-1}$ error is shown in red.  The green dashed line marks the dichroic separation at 5600\text{\AA}.  The shape of the spectrum becomes less reliable shortward of 3600\text{\AA} (1060\text{\AA} restframe) due to systematics in flat-fielding.  Longward of 7000\text{\AA} (2060\text{\AA} restframe), the spectrum becomes more sky-dominated and the smoothed spectrum has been manually clipped.  The yellow dashed box is the zoomed-in region shown in the lower plot.  \textit{Bottom:} The same spectrum zoomed-in, smoothed and redshifted to $z=2.399$.  A weak Ly-$\alpha$ absorber LBG composite spectrum (see text) is overlaid in blue and scaled arbitrarily to emphasize the alignment of features.  A subset of the features used to constrain the redshift is illustrated with dashed blue lines and labeled.}
    \label{fig:16adga_spec}
\end{figure*}

We collected the Keck LRIS spectrum of HSC16adga on MJD~57751.09.  The positioning of the spectral slit is mapped in Figure~\ref{fig:hst}.  From the light curve in Figure~\ref{fig:16adga_phot} the spectrum appears to have been taken near peak in the restframe-UV.  The spectrum, composed of 6 exposures (7200s blue, 7074s red), was reduced using standard \texttt{IRAF}\footnote{IRAF is distributed by the National Optical Astronomy Observatories, which are operated by the Association of Universities for Research in Astronomy, Inc., under cooperative agreement with the National Science Foundation.} procedures.  The combined blue and red-side 1-D spectrum is shown in Figure~\ref{fig:16adga_spec}. 

The spectrum includes a significant host galaxy flux contribution from which we derive a spectroscopic redshift.  The photometry of the host is suggestive of a high redshift starforming galaxy such as a LBG.  The accepted method of spectroscopic redshift determination of high redshift LBGs is to infer an initial coarse redshift from visual confirmation of \lya, the \lya forest decrement and strong interstellar medium (ISM) absorption lines, and to then constrain the redshift by comparison to LBG composite spectra with full FUV line lists and consideration of the behavior of the lines as a result of outflows (i.e., the ubiquitous presence of $\sim$100--200km s$^{-1}$ blueshifted ISM absorption line profiles and the P-Cygni-like profile of the \lya line; \citealp{steidel96,Steidel98,steidel03,steidel04,pettini00,pettini01,shapley03,adelberger03,cooke06lbgs,jones12}).  

In the reduced spectrum of HSC16adga we identify a strong absorption feature at $\sim$4100\text{\AA}, observer-frame, with a slight flux break in the continuum at shorter wavelengths, which we attribute to Ly-$\alpha$ absorption and the Ly-$\alpha$ forest, respectively.  For comparison throughout we use the LBG composite spectra of \citet{shapley03}.  From Figure~\ref{fig:16adga_spec} several features are apparent (e.g. Si\textsc{ii} $\lambda$1260, O\textsc{i} $\lambda$1302, C\textsc{ii} $\lambda$1335, Si\textsc{iv} $\lambda$1394, O\textsc{iv]} $\lambda$1403), indicating a redshift of $z = 2.399\pm0.004$, consistent with both the COSMOS2015 photometric redshift estimate and \texttt{MIZUKI} redshift probability distribution.  The error is given in terms of the spectral resolution of LRIS, doubled to account for a 2-pixel smoothing algorithm applied to clarify absorption features.

At present spectroscopic cross-correlation is ineffective for the positive determination of precise redshifts of high redshift LBGs.  One reason for this is because none of the identifiable FUV features of LBGs are at rest with respect to the redshift of the LBG system itself \citep{steidel03}.  Still, cross-correlation can be used to quickly rule out the possibility of an alternative low redshift fit.

We perform a simple cross-correlation of the spectrum with the Manual and Automatic Redshifting Software (\texttt{MARZ}; \citealp{hinton16}).  \texttt{MARZ} cross-correlates input spectra with a library of galaxy templates and outputs the five most likely fits.  Each fit includes a likelihood score from 1--4 representing respectively a likelihood of less than 50\%, 50\%--90\%, 90\%--99\% and greater than 99\%.  The \texttt{MARZ} top five redshift fits to the spectrum of HSC16adga range from $0.2845<z<2.42$ with each fit receiving a score of 1, less than 50\% likely.

As an independent check for the visual redshift estimates, we perform a coarse fit on the observed spectra with a library of spectral templates to infer their redshifts.  The strong night sky lines are not always subtracted perfectly in spectroscopic observations, and the sky residuals are the main source of systematic uncertainties in observed spectra, which are not captured by noise spectra.  To suppress such systematics, we bin the spectra in 200\AA windows (although in practice the inferred probability distributions are not very sensitive to the choice of bin size) by first clipping the top and bottom 15\% of the flux distribution in each bin and then taking the weighted average of the remaining data points.  The clipping here is intended to largely eliminate the systematics.  We then feed the binned fluxes to the photo-$z$ code, \texttt{MIZUKI}, to infer narrow-band redshift probability distributions to compliment the photometric redshifts derived from broad-band photometry (see M19).  Only galaxy templates are used in the fitting, and the supernova contribution to the spectrum is a source of uncertainty.

\begin{figure}
\begin{center}
\includegraphics[height=6.7cm]{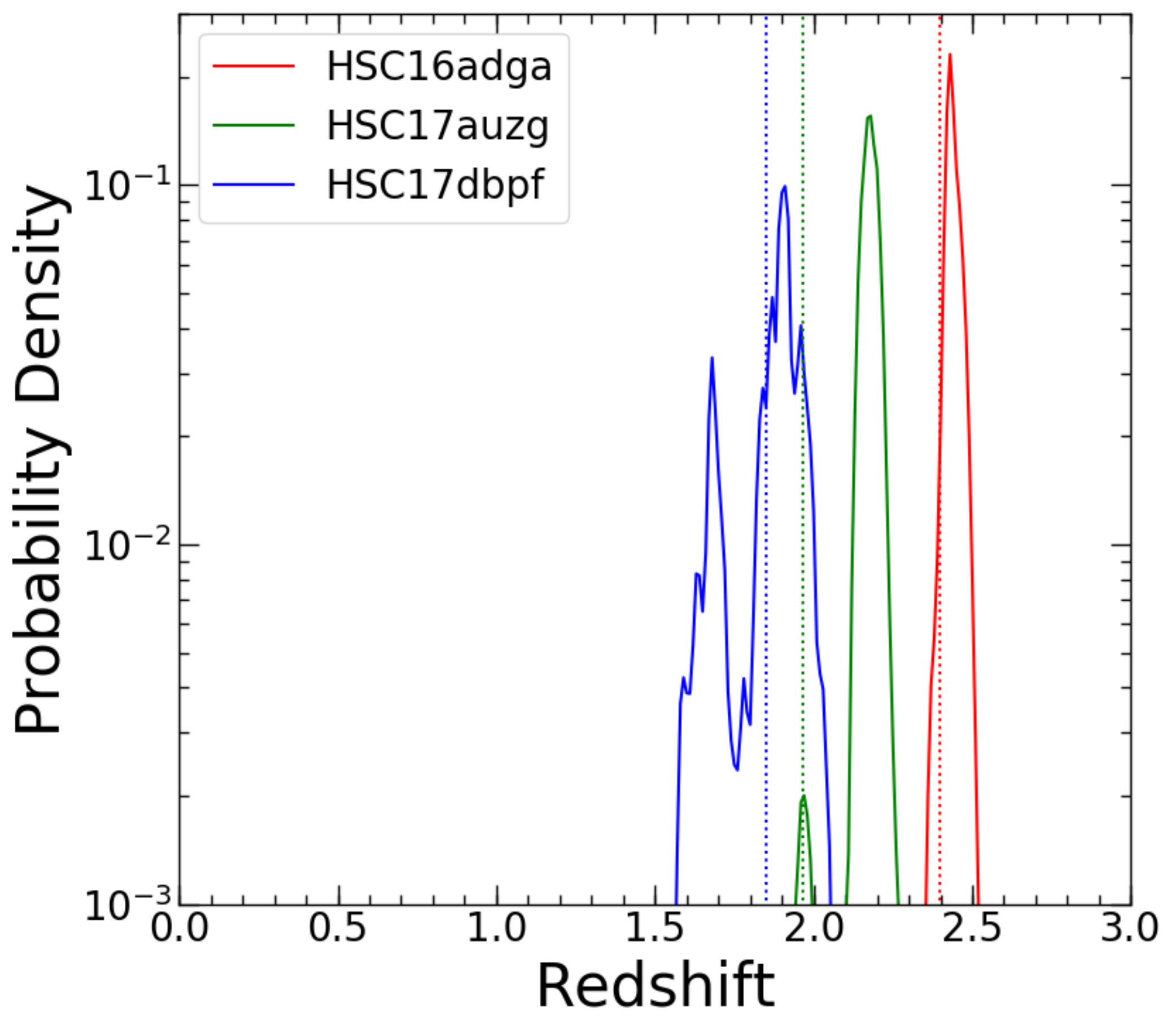}
\end{center}
\caption{\texttt{MIZUKI} narrow-band redshift probability distributions for each event.  The \texttt{MIZUKI} broad-band photometric redshift probability distributions are presented in identical fashion in M19.}
\label{fig:mizuki_2}
\end{figure}

The \texttt{MIZUKI} narrow-band redshift estimate of the spectrum of HSC16adga is shown in Figure~\ref{fig:mizuki_2}.  Compared to the broad-band estimate, the probability of the spectroscopic redshift is higher because several alternative high redshift probabilities have been eliminated (see M19).

\subsection{HSC17auzg}

HSC17auzg (SN-2016jhn) was first detected by SHIZUCA on MJD~57745 at coordinates (RA, Dec) = (09:59:00.42, +02:14:20.8).  The transient is detected in a star-forming host with a COSMOS photo-$z$ of $1.65^{+0.06}_{-0.08}$ and a \texttt{MIZUKI} photo-$z$ peaking at $1.78$.  It is observed offset by 0\farcs78$\pm$0\farcs051 ($\sim$6kpc) from the host galaxy flux centroid, helping to rule out an AGN or a TDE.  The light curve of HSC17auzg is shown in Figure~\ref{fig:16adga_phot}.

\begin{figure*}
	\includegraphics[width=\textwidth]{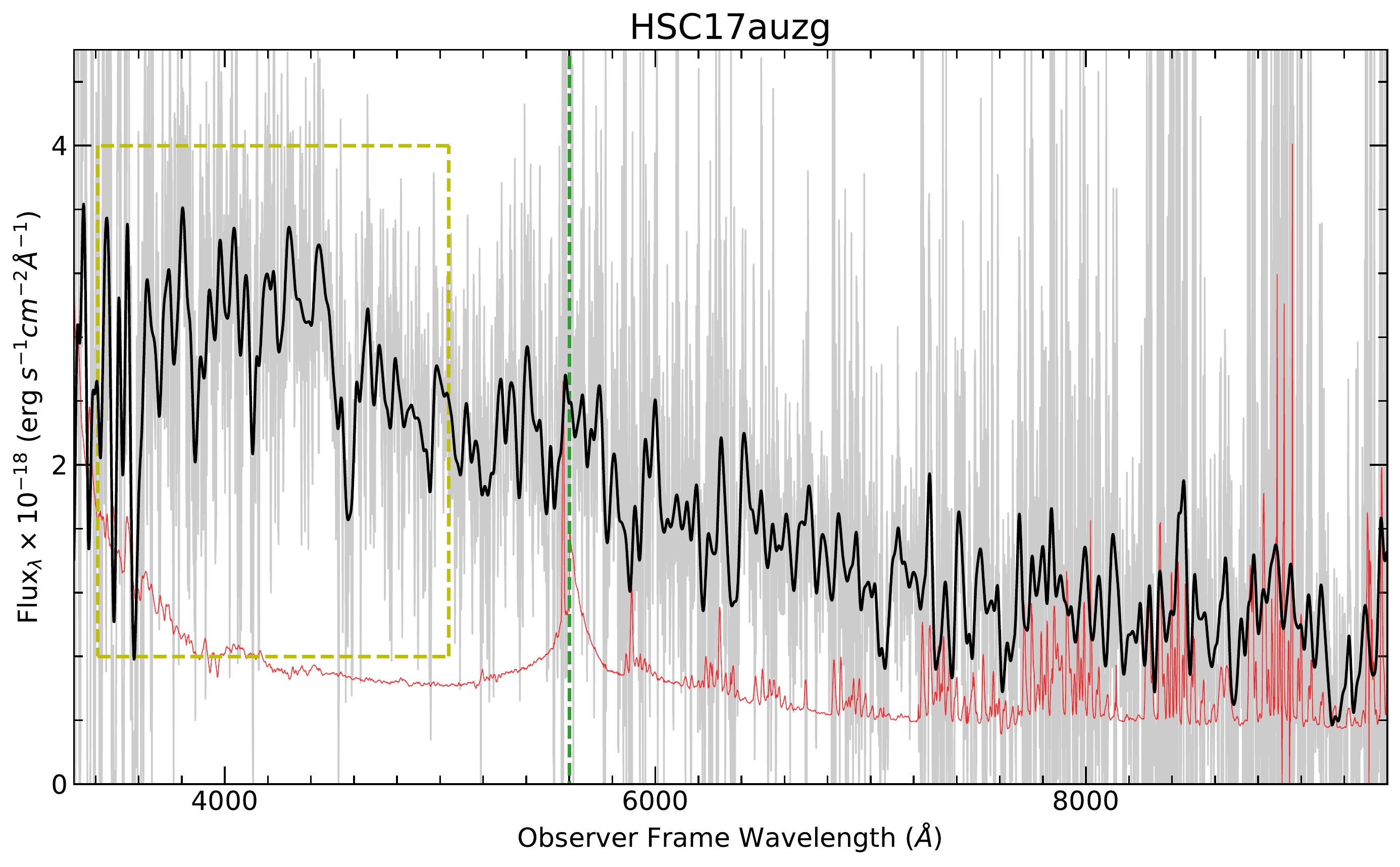}
    \includegraphics[width=\textwidth]{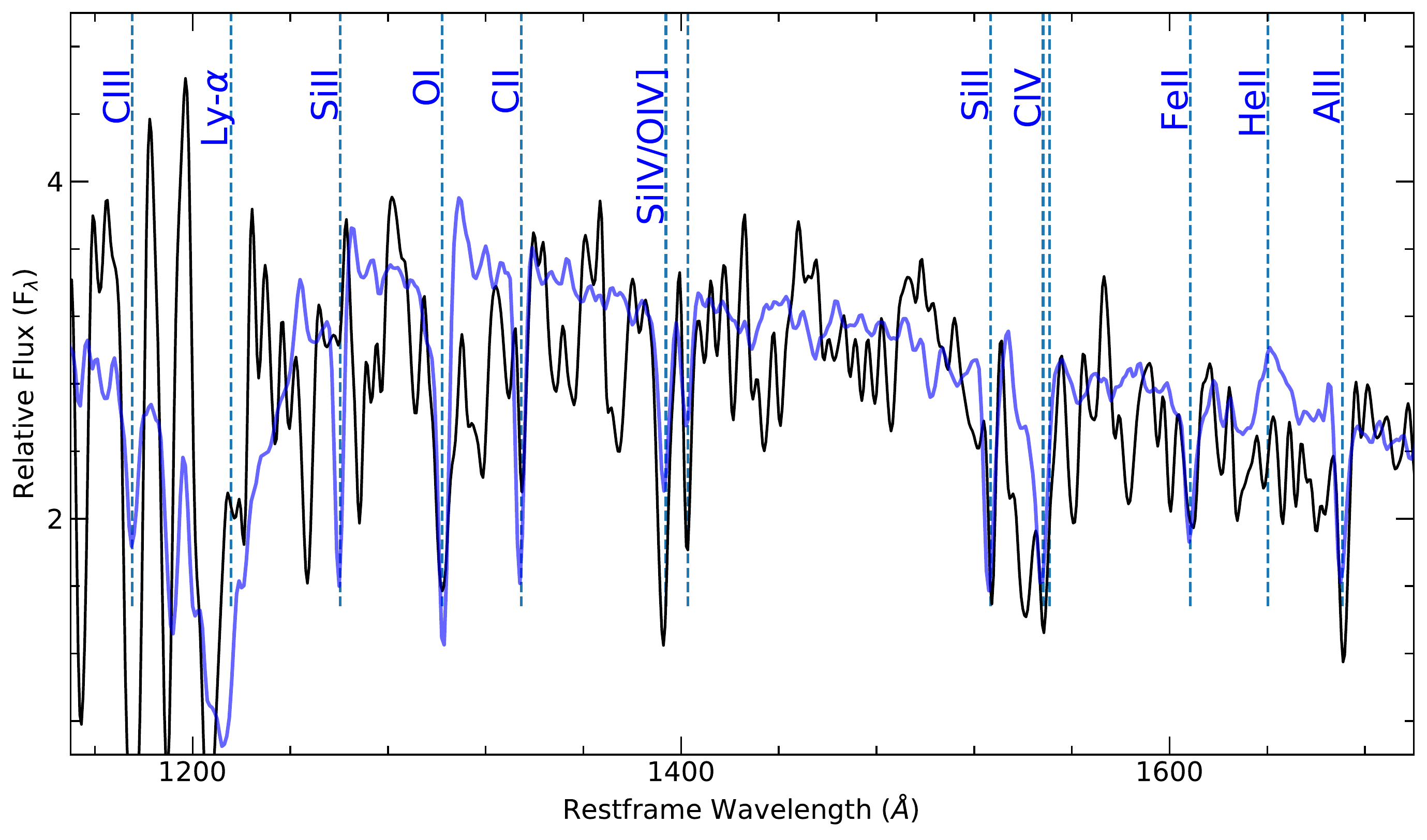}
   \caption{\textit{Top:} Flux-calibrated observer-frame 1-D spectrum of HSC17auzg, similar to Figure~\ref{fig:16adga_spec}.  The shape of the spectrum becomes less reliable shortward of 3600\text{\AA} (1215\text{\AA} restframe), due to systematics in flat-fielding.  Longward of 7000\text{\AA} (2360\text{\AA} restframe) the spectrum becomes more sky-dominated. \textit{Bottom:} The same spectrum zoomed-in, smoothed and redshifted to $z=1.965$, similar to Figure~\ref{fig:16adga_spec}.  A strong Ly-$\alpha$ absorber LBG composite spectrum (see text) is overlaid in blue and scaled arbitrarily to emphasize the alignment of features.  A subset of the features used to constrain the redshift is illustrated.}
    \label{fig:17auzg_spec}
\end{figure*}

We collected the Keck LRIS spectrum of HSC17auzg on MJD~57835.80.  The positioning of the spectral slit is mapped in Figure~\ref{fig:hst}.  From the light curve in Figure~\ref{fig:16adga_phot} we estimate that the spectrum was taken some days after the restframe-UV peak.  The spectrum, composed of 7 exposures (8400s blue, 8253s red), was reduced using standard \texttt{IRAF} procedures.  The 1-D spectrum is shown in Figure~\ref{fig:17auzg_spec}.

We identify a strong absorption feature at $\sim$3500\text{\AA}, observer-frame, which we attribute to Ly-$\alpha$.  From this we make a first approximation of the redshift whereby several key ISM absorption features of LBGs become apparent (e.g. O\textsc{i} $\lambda$1302, Si\textsc{iv} $\lambda$1394, O\textsc{iv]} $\lambda$1403, Si\textsc{ii} $\lambda$1527, C\textsc{iv} $\lambda\lambda$1548, 1551, Al\textsc{ii} $\lambda$1671), indicating a redshift of $z = 1.965\pm0.004$ (see Figure~\ref{fig:17auzg_spec}).  This is significantly higher than the COSMOS2015 photometric redshift estimate but is represented in the \texttt{MIZUKI} redshift probability distribution.

The \texttt{MARZ} five most likely cross-correlation redshift fits to the spectrum of HSC17auzg range from $0.428<z<3.95$ with each fit receiving a score of 1, less than 50\% likely.

We produce a \texttt{MIZUKI} narrow-band redshift estimate for HSC17auzg (see Figure~\ref{fig:mizuki_2}).  Compared to the broad-band estimate (see M19) the probability of the spectroscopic redshift is similar but its proximity to a probability peak is significantly improved.  There is a more dominant probability peak, however spectroscopic feature alignment occurs only at the redshift of the secondary probability peak.

\subsection{HSC17dbpf}

HSC17dbpf (SN-2017fei) was first detected by SHIZUCA on MJD~57816 at coordinates (RA, Dec) = (09:58:33.42, +01:59:29.7).  The transient is detected in a star-forming host with a COSMOS photo-$z$ of $2.25^{+0.08}_{-0.53}$ and a \texttt{MIZUKI} photo-$z$ peaking at $1.58$.  It is observed offset by 0\farcs58$\pm$0\farcs052 ($\sim$5kpc) from the host galaxy flux centroid, helping to rule out an AGN or a TDE.  The light curve of HSC17dbpf is shown in Figure~\ref{fig:16adga_phot}.

\begin{figure*}
	\includegraphics[width=\textwidth]{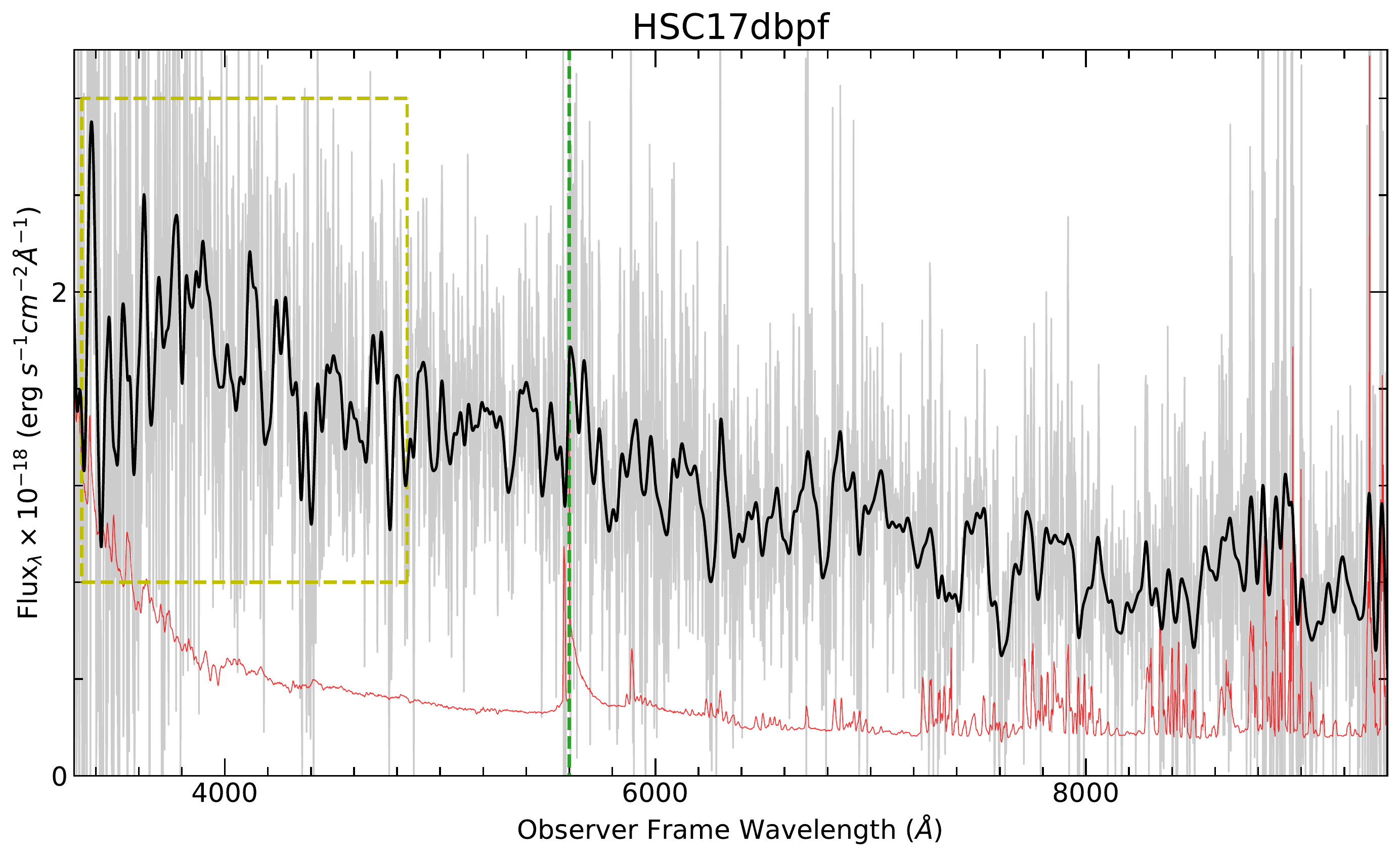}
    \includegraphics[width=\textwidth]{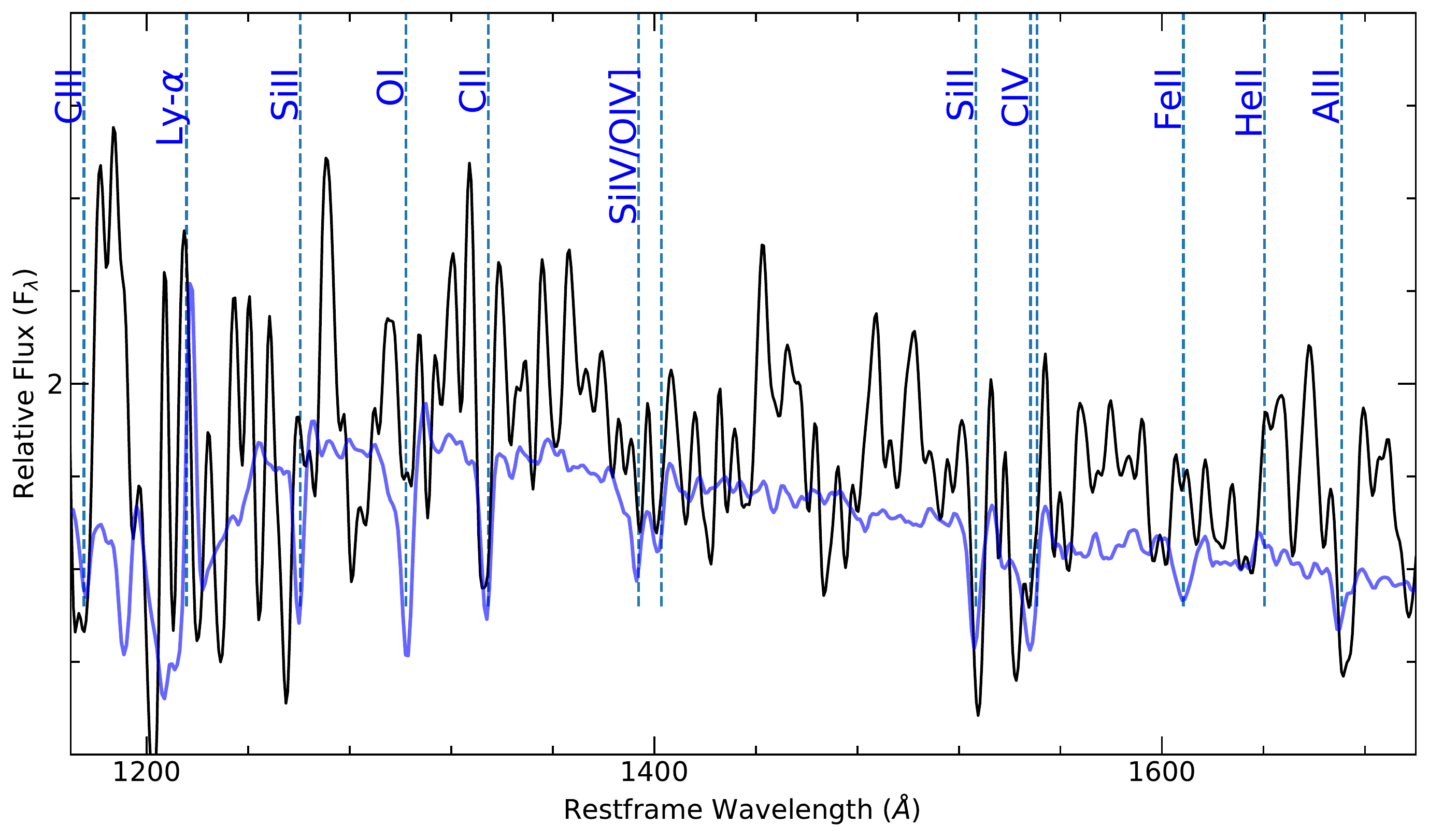}
   \caption{\textit{Top:} Flux-calibrated observer-frame 1-D spectrum of HSC17dbpf, similar to Figure~\ref{fig:16adga_spec}.  The shape of the spectrum becomes less reliable shortward of 3600\text{\AA} (1260\text{\AA} restframe) due to systematics in flat-fielding.  Longward of 7000\text{\AA} (2455\text{\AA} restframe), the spectrum become more sky-dominated. \textit{Bottom:} The same spectrum zoomed-in, smoothed and redshifted to $z=1.851$, similar to Figure~\ref{fig:16adga_spec}.  A weak Ly-$\alpha$ absorber LBG composite spectrum (see text) is overlaid in blue and scaled arbitrarily to emphasize the alignment of features.  A subset of the features used to constrain the redshift is illustrated.}
    \label{fig:17dbpf_spec}
\end{figure*}

We collected two Keck LRIS spectra of HSC17dbpf on MJD~57835.93 and MJD~57836.84.  The positioning of the spectral slit is mapped in Figure~\ref{fig:hst}.  The light curve in Figure~\ref{fig:16adga_phot} indicates that the event evolved very quickly, making the timing of the restframe-UV peak difficult to estimate.  But it appears that the spectra were taken within a few days of this peak.  The spectra (see Figure~\ref{fig:17dbpf_spec}) were collected over 2 consecutive nights which we treat as a single epoch composed of 5 exposures (6000s blue, 5895s red). 

We attribute the strong absorption feature at $\sim$3400\text{\AA}, observer-frame, to Ly-$\alpha$.  From this we make a first approximation of the redshift whereby several key ISM absorption features of LBGs become apparent (e.g. C\textsc{ii} $\lambda$1335, Si\textsc{iv} $\lambda$1394, O\textsc{iv]} $\lambda$1403, Si\textsc{ii} $\lambda$1527, C\textsc{iv} $\lambda$$\lambda$1548, 1551), indicating a redshift of $z=1.851\pm0.004$ (see Figure~\ref{fig:17dbpf_spec}).  This is consistent with the COSMOS2015 photometric redshift estimate but higher than predicted by the \texttt{MIZUKI} redshift probability distribution.

The \texttt{MARZ} five most likely cross-correlation redshift fits to the spectrum of HSC17dbpf range from $0.0155<z<4.67$ with each fit receiving a score of 1, less than 50\% likely.

We produce a \texttt{MIZUKI} narrow-band redshift estimate for HSC17dbpf (see Figure~\ref{fig:mizuki_2}).  Compared to the broad-band estimate which is inconsistent with the spectroscopic redshift (see M19), the probable redshift range is similar but encompasses higher values including the spectroscopic redshift.

\section{DISCUSSION}

\subsection{Nature}

The high redshift supernova nature of each event is well-supported.  The spectroscopic redshifts are precisely constrained and consistent with the \texttt{MIZUKI} narrow-band redshifts which offer no low redshift probability, and cross-correlation with \texttt{MARZ} produces no reliable low redshift alternative. Furthermore, when the measured redshifts are applied to the transient photometry, M19 find low redshift supernovae with comparable UV light curves in each case.

\begin{figure*}
\includegraphics[width=\textwidth]{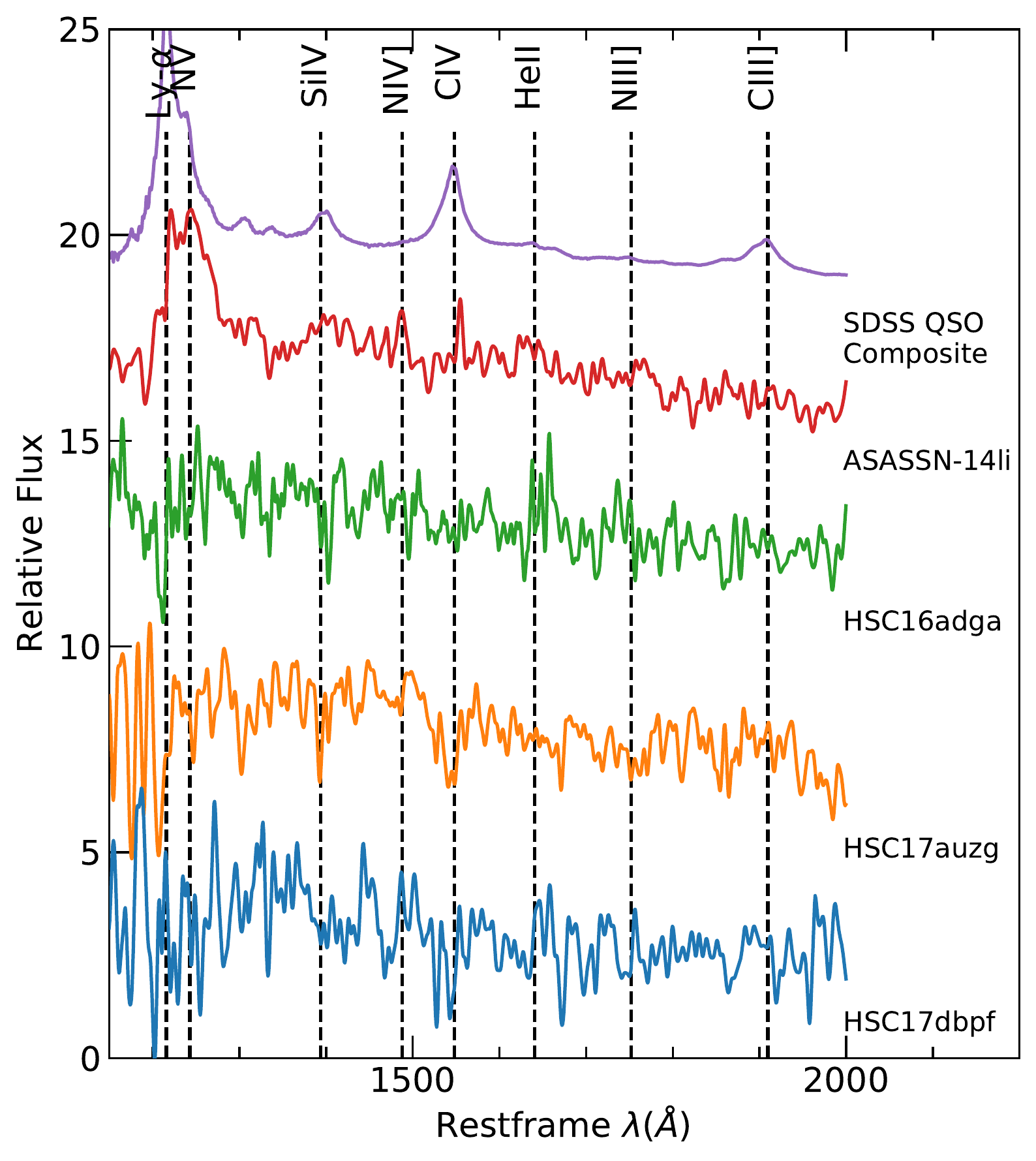}   \caption{All three supernova spectra including their host galaxy contributions.  Also presented are a QSO composite and the TDE, ASASSN-14li.  We have increased the noise in the spectrum of the TDE to approximate that of our observations, and have applied a Ly-$\alpha$ deforestation factor to the TDE and the QSO at wavelengths shortward of 1216\text{\AA} consistent with a redshift of $z\sim2$.  The labeled transitions are emission features observed in ASASSN-14li by \citet{cenko1614li}, most of which are also commonly seen in AGN (as is evident from the QSO composite).}
    \label{fig:agnlike}
  \end{figure*}

Although TDEs  can resemble supernovae in their light curve behavior, these are highly centralized events and our candidates are all significantly offset from their host galaxy flux centroids (see Figure~\ref{fig:hst}).  Thus it is unlikely that the candidates are of this type.  In Figure~\ref{fig:agnlike} we compare the observed spectra to the FUV spectrum of the TDE, ASASSN-14li \citep{cenko1614li}.  We include a Sloan Digital Sky Survey (SDSS) quasi-stellar object (QSO, equivalent to AGN) composite spectrum \citep{vanden01sdssqso} to which the TDE is similar.  The lack of any strong N\textsc{v} emission or any other nitrogen emission in the observed spectra is a notable difference to that expected of a TDE.

The SLSN class of supernovae was first defined as a luminosity class that includes all supernovae exceeding an absolute magnitude of M $=-21$ in any band \citep{galyam12}.  However, currently SLSNe are identified and grouped spectroscopically and comprise event with a continuum of peak magnitudes which extends below the M $=-21$ threshold \citep{nicholl17magnetar,moriya2018slsne,quimby18slsnIspec}.  

The k-corrected peak absolute magnitudes of the presented SHIZUCA photometric candidates (see M19) approach but do not meet the strict photometric criteria of SLSNe set by \citet{galyam12}.  However the over-luminous aspect of each of the candidates in the FUV, while difficult to reconcile with ordinary supernovae, is representative of SLSNe \citep{quimby11,yan1716apd,Yan17egm}.  The intermediate FWHM seeing and light cirrus during spectroscopic follow-up resulted in the candidate spectra having lower S/N than anticipated and prevents positive supernova type-determinations.  Thus as implied throughout we identify these transients as probable SLSNe.  Late-time follow-up of the candidates could enable supernova type-determinations by providing more precise host galaxy subtraction templates and revealing any late-time circumstellar emission features as seen in some SLSNe \citep{fox15,yan17lateh}.    

\subsection{Host Subtraction}

We compare our host-subtracted spectra to some of the few FUV spectra of early-time SLSNe and UV-bright SNe in an effort to discern the spectroscopic type of each supernova.  While unsuccessful in this regard, the comparisons still offer useful information.

Because we do not have spectra of the host galaxies alone, we perform a ``pseudo-host'' subtraction on each spectrum to enable our analysis.  We assume a LBG-like host for each supernova, consistent with the templates fit with \texttt{MIZUKI} to the broad-band photometry (see M19) and with the hosts of SLSNe in general.  Analyses of a large sample of LBGs \citep{shapley03,cooke09snIIns} have shown that the characteristics and colors of a LBG spectrum are strongly correlated with the equivalent width of the Ly-$\alpha$ feature.  \citet{shapley03} divided their LBG sample into quartiles based on this equivalent width and subsequently constructed composite spectra for each.  The quartiles are referred to here as strong Ly-$\alpha$ absorbers, weak Ly-$\alpha$ absorbers, weak Ly-$\alpha$ emitters and strong Ly-$\alpha$ emitters.  We include these composites in our pseudo-host constructions using the estimated Ly-$\alpha$ absorption or emission strengths of the hosts to select the specific composite for use in each subtraction.  

The wavelength ranges of the spectra include the HSC-\textit{gri} bands.  To construct the pseudo-hosts over this whole range we use the LBG composite spectra from \mbox{912\text{\AA}--2000 \text{\AA}} (their full extent) and starburst (SB) templates from \citet{calzetti94} at wavelengths beyond 2000\text{\AA}.  This extension is valid because LBGs are highly star-forming galaxies.  Ly-$\alpha$ absorbers have redder UV continua compared to Ly-$\alpha$ emitters \citep{shapley03,cooke09snIIns}.  We therefore use the SB2 template (0.10 $<$ E(B-V) $<$ 0.21) to complete a pseudo-host constructed with a Ly-$\alpha$ absorber LBG composite, and we use the SB1 template (E(B-V) $<$ 0.10) to complete a pseudo-host constructed with a Ly-$\alpha$ emitter LBG composite.

To accurately scale the pseudo-hosts we calculate the ratio of host-flux to supernova-flux in each spectrum using the associated HSC photometry as measured most closely in time to our observations (see M19).  The high redshift hosts are near-point sources and the bulk of the flux enters the spectroscopic slit in each case.  But each slit is centered on the supernova, thus some of the host flux may not be included (see Figure~\ref{fig:hst}).  When necessary we apply a geometrical correction to account for any portion of the host not within the spectroscopic slit.  

\subsection{SLSNe-II}

Type-II supernovae (SNe-II) are defined as those that exhibit hydrogen lines in spectra, typically H-$\alpha$ \citep{filippenko97}, though they can also be confirmed with Ly-$\alpha$ emission in FUV spectra \citep{fransson02,fransson05,fransson14}.  LBGs, which are highly star-forming galaxies, are a common host type for high redshift supernovae \citep{cooke09snIIns,cooke12hiz}.  But they are also often strong Ly-$\alpha$ absorbers or emitters, which complicates the task of identifying supernova Ly-$\alpha$ emission.

Observed Ly-$\alpha$ emission in a high redshift supernova/host spectrum can still be used as a positive identifier of a SNe-II if it can be confidently attributed to the supernova.  In cases where the host and supernova are spatially separated on the sky such that their profiles are distinguishable, the source or sources of Ly-$\alpha$ emission can be observed directly.  If this is not the case the source of observed Ly-$\alpha$ emission can still be inferred from the characteristics of the feature.  Ly-$\alpha$ emission from a supernova originates in the ejecta or a circumstellar medium (CSM) or both. Ly-$\alpha$ emission from ejecta is broadened according to the ejecta velocity.  Ly-$\alpha$ from a cold, slow-moving CSM is observed as a narrow feature with a blue velocity offset from the redshift of the host.  In cases of LBGs with Ly-$\alpha$ in emission, the feature arises as Ly-$\alpha$ emission from the LBG is back-scattered off receding gas outflows and passes back through the LBG off-resonance avoiding absorption.  Consequently the (narrow) feature is typically redshifted from the host by up to 1000km~s$^{-1}$ or more \citep{shapley03}, thus distinguishing it from any relatively blue-shifted supernova Ly-$\alpha$ emission.

High redshift supernova Ly-$\alpha$ emission can also be largely or completely absorbed by neutral hydrogen (H\textsc{i}) in the host interstellar medium, its circumgalactic medium, and the intervening intergalactic medium.  So the absence of observable supernova Ly-$\alpha$ emission does not necessarily imply the absence of hydrogen integral to the explosion scenario.  In cases where the lack of Ly-$\alpha$ emission is ambiguous, the presence of certain features unattributable to the host alone may be another reliable way to identify a SLSN-II from its FUV spectrum.  For example, most SLSNe-II are spectroscopically classified as SNe-IIn based on the presence of narrow hydrogen lines in emission \citep{06gy_ofek07,smith0706gyobs,06tf_smith08,08fz_drake10,08am_chatzopoulos11}.  Studies by \citet{fransson05,fransson14} have shown that SN-IIn FUV spectra can have signature strong emission features such as O\textsc{i} $\lambda$1300, N\textsc{iv} $\lambda$1486 and Mg\textsc{ii} $\lambda$2800.  These features could conceivably be used to confirm a SLSN-II classification \citep{cooke09snIIns}.  Magnesium emission would be a particularly useful identifying feature of SLSNe-II as it has been observed to persist strongly to very late times in SNe-IIn \citep{fransson05}.

\begin{figure*}
\includegraphics[width=\textwidth]{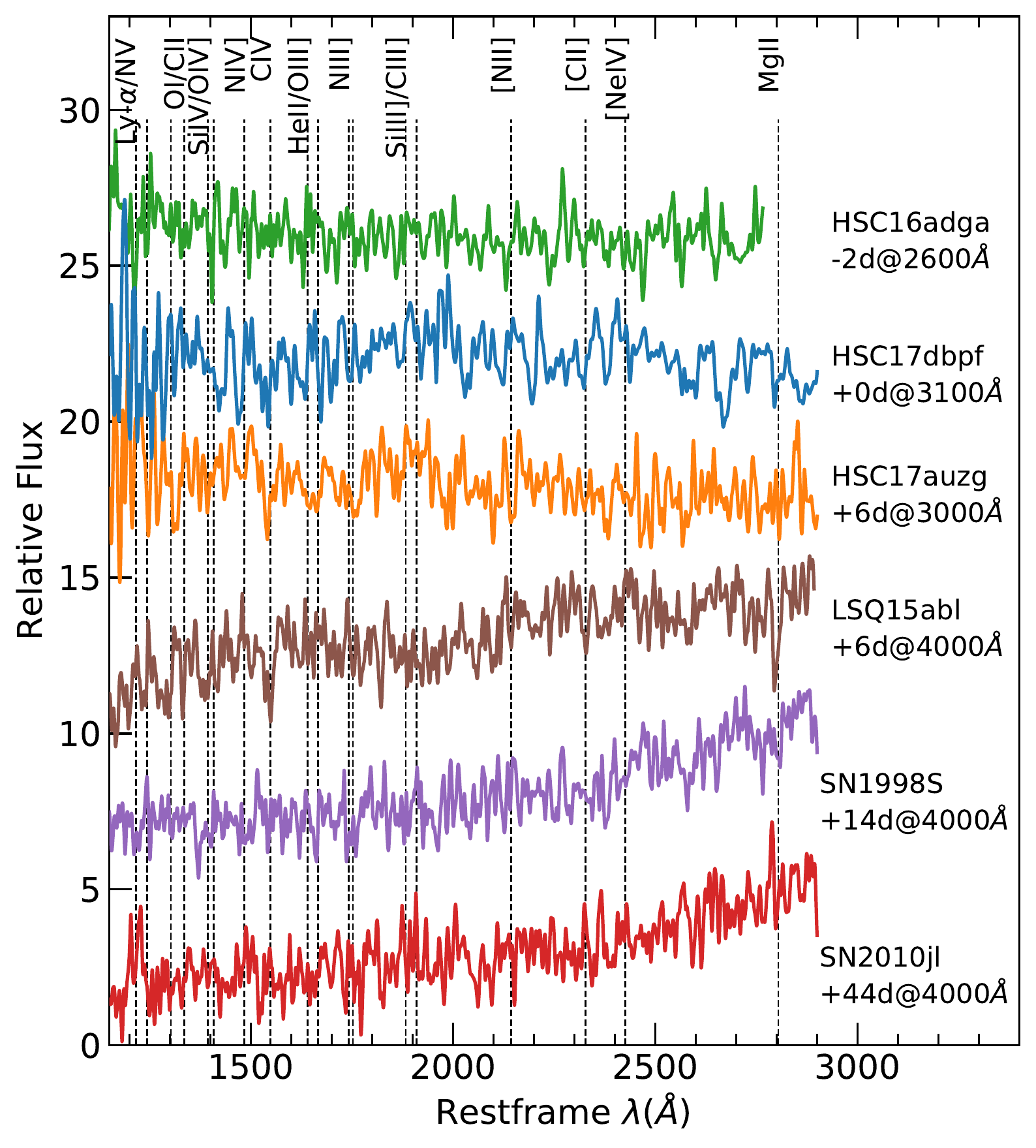}   \caption{The three pseudo-host-subtracted supernova spectra (see text) along with the SLSN-II, LSQ15abl, and 2 UV-bright SNe-IIn.  The spectra are labeled along with their approximate phase in days since peak at the associated restframe wavelength.  We have increased the noise in the comparison spectra to approximate that of our observations, and have applied a Ly-$\alpha$ forest effect for a redshift of $z\sim2$ at wavelengths shortward of 1216\text{\AA}.  The labeled transitions are emission features identified in SNe-IIn by \citet{fransson05,fransson14}.}
    \label{fig:sl2like}
  \end{figure*}

In Figure~\ref{fig:sl2like} we compare our pseudo-host-subtracted spectra to the early-time FUV spectrum of the low redshift SLSN-II, LSQ15abl \citep{lsq15abl}, and the earliest FUV spectra of the UV-luminous SNe-IIn, SN1998S \citep{fransson05} and SN2010jl \citep{fransson14}.  None of our supernovae exhibit obvious Ly-$\alpha$ emission.  It is tempting to attribute the blunted emission-like feature at $\sim$1190\text{\AA} in the spectrum of HSC17dbpf to a component of slightly blue-shifted Ly-$\alpha$ emission, but this feature arises at a wavelength at which the shape of the spectrum is unreliable due to systematics in flat-fielding.  It is worth noting that in the noise-added spectra being used for comparison, Ly-$\alpha$ emission is likewise not obvious.  We know that these events do exhibit Ly-$\alpha$ emission, so this comparison illustrates the baseline S/N required of a high redshift supernova spectrum in order to rule out at least any unmolested Ly-$\alpha$ emission.  The coarse blackbody shapes of our spectra are bluer than the comparison spectra, but this is not surprising given their earlier collection times.  Beyond these observations, the spectra are simply of too low S/N to say anything about the more subtle emission features anticipated in the FUV of SLSNe-II.

\subsection{SLSNe-I}

The spectra of SLSNe-I are uniquely conspicuous, displaying a number of characteristic broad hump-like features throughout the optical and UV \citep{quimby11,howell13hizslsne,nicholl15Ics,yan1716apd,Yan17egm}.  The number of spectroscopically observed SLSNe-I is increasing rapidly and the nature of these features is being clarified \citep{mazzali16,quimby18slsnIspec,lunnan18}. 

Our understanding of the UV features of SLSNe-I is also improving thanks in large part to spectroscopic studies of high redshift examples \citep{berger1211bam,howell13hizslsne,pan15e2mlf,smith18c2nm}.  From what has been observed to date, SLSNe-I typically exhibit broad absorption troughs centered around $\sim$1930\text{\AA}, 2200\text{\AA}, 2400\text{\AA} and 2660\text{\AA} \citep{smith18c2nm}.  Spectral synthesis models have attributed these features to blended transitions of iron, cobalt, carbon, titanium, silicon and magnesium \citep{mazzali16}.  Prominent features at still shorter wavelengths (e.g., 1400\AA, 1700\AA) have been observed in spectra of high redshift SLSNe-I for which UV coverage extends sufficiently far \citep{pan15e2mlf}, as well as in low redshift SLSNe-I with HST FUV spectral coverage \citep{yan1716apd,Yan17egm}.  But the consistency of these features among such events is insufficient to draw conclusions about their normalcy at this time.

\begin{figure*}
\includegraphics[width=\textwidth]{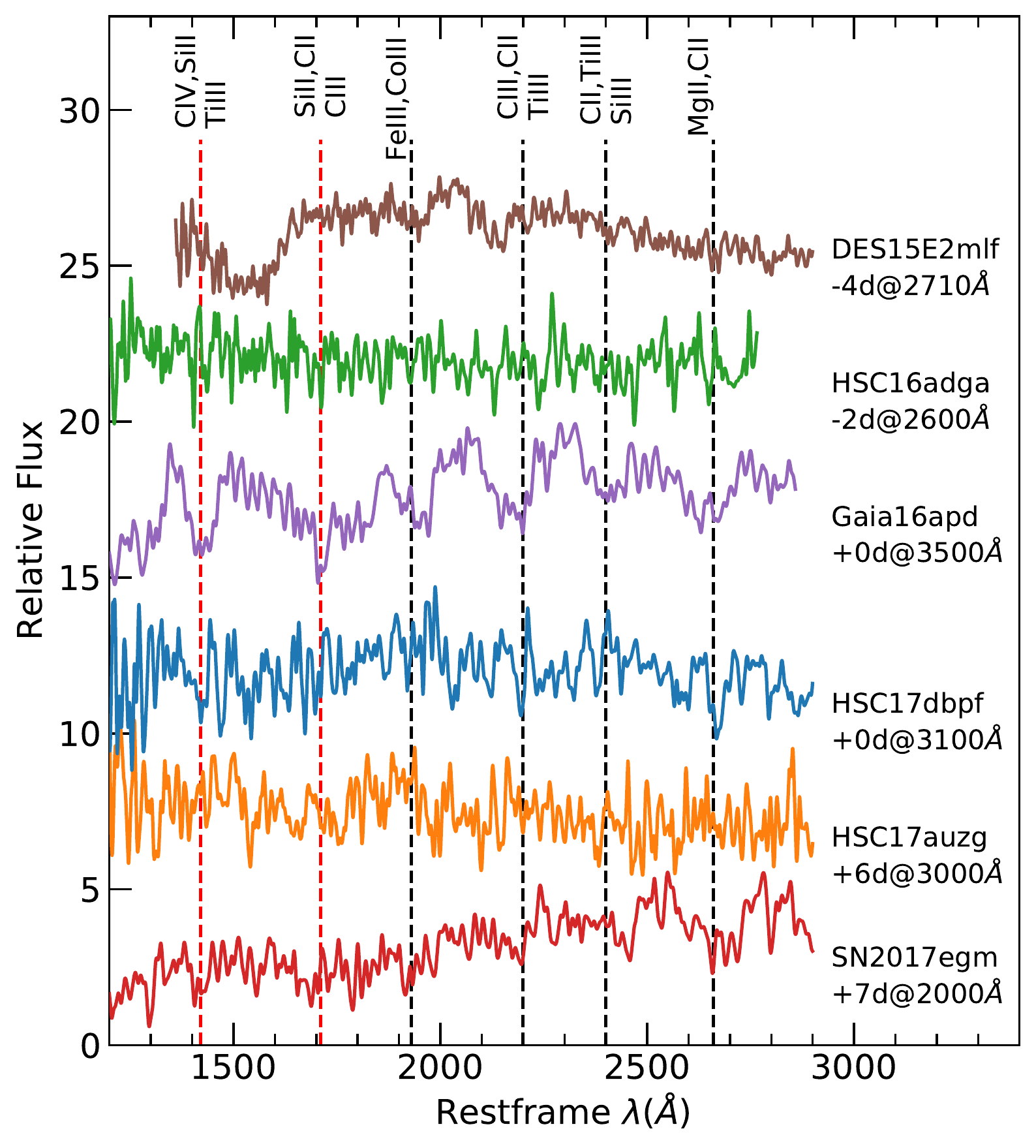}   \caption{The three pseudo-host-subtracted supernova spectra (see text) along with the high redshift SLSN-I, DES15E2mlf, and the two low redshift SLSNe-I.  The spectra are labeled along with their approximate phase in days since peak at the associated restframe wavelength.  We have increased the noise of the comparison spectra to approximate that of our observations, and have applied a Ly-$\alpha$ forest effect for a redshift of $z\sim2$ at wavelengths shortward of 1216\text{\AA}.  Transitions reproduced in the spectral models of \cite{mazzali16} are marked with black dotted lines, while those marked with red dotted lines are derived by \citet{Yan17egm} with a model fitting method.}
    \label{fig:sl1like}
  \end{figure*}

In Figure~\ref{fig:sl1like} we compare our pseudo-host-subtracted spectra to an extensive FUV spectrum of the high redshift SLSN-I, DES15E2mlf \citep{pan15e2mlf}, and the two low redshift SLSNe-I with FUV spectral coverage, Gaia16apd \citep{yan1716apd} and SN2017egm \citep{Yan17egm}.  In the noise-added comparison spectra the absorption troughs are significantly dampened, but most are still discernible.  Differences in the strengths, widths and central wavelengths of individual troughs are also apparent.  The longer wavelength features are not observed to any significance in our high redshift sample, though at these wavelengths our spectra are all sky-dominated.  There seem to be some similarities between the comparison spectra and the spectra of HSC17auzg and HSC17dbpf at short wavelengths (e.g. the absorption dip at $\sim1400\text{\AA}$, the general trends of the spectra from $\sim1700-2000\text{\AA}$).  Positive SLSN-I classifications can be made with optical spectra by performing a $\chi^2$ minimization of model spectra created by scaling, adding, and then redshifting supernova and host galaxy templates \citep{howell05,quimby18slsnIspec}. We attempt to quantify the significance of any similarities between our spectra and the FUV spectra of SLSNe-I by the same method.  However with so few FUV spectra of SLSNe-I available for comparison the significance of the presence or lack of features in the FUV spectra of our targets
cannot be quantified and our results are inconclusive. 

\section{CONCLUSION}
With this SHIZUCA pilot spectroscopic program we have demonstrated the feasibility of spectroscopic follow-up of high redshift SLSNe, testing the current limits of high redshift transient spectral analysis.  The high \'entendu and dense cadence of the HSC-SSP combined with the detection efficiency of SHIZUCA can generate high redshift SLSN candidates at a sufficient rate to statistically enable follow-up of several targets near maximum light at any time during an active observation campaign.  Keck LRIS provides the best facility for spectroscopic follow-up of $z\sim2$ supernovae in terms of light-gathering power combined with blue sensitivity.  

We have reported on FUV spectroscopic follow-up of three probable SLSNe at $z=1.851,$ 1.965 and 2.399.  The spectra are extracted from $\sim$2 hour exposures taken under sub-optimal seeing and weather conditions.  The S/N of the reduced spectra are relatively low, and flux measurements become unreliable at wavelengths shortward of 3600\text{\AA}, observer-frame.  Thus in LRIS spectra of similar S/N, moderate-to-weak Ly-$\alpha$ features, perhaps dampened by H\textsc{i} absorption, cannot be confirmed or ruled out at $z\lesssim2$.  Such analyses can be extended to shorter wavelengths and lower redshifts using spectra with higher S/N or exhibiting strong Ly-$\alpha$ emission.

Our analysis of the pseudo-host-subtracted supernova spectra is inconclusive regarding the spectroscopic type of each event.  We estimate that doubling the S/N of the spectra presented here would enable supernova type-confirmations of future events by substantiating any strong emission features of SLSNe-II or the broad absorption troughs of SLSNe-I.  By observing under optimal seeing conditions, employing longer exposures, and perhaps taking advantage of brighter than average targets, this S/N is achievable.

Future follow-up is merited to ascertain if any of the supernovae exhibit late-time strong emission (e.g., Ly-$\alpha$, Mg\textsc{ii}) from which type-information can be inferred.  In addition, late-time spectra of the hosts with little-to-no supernova contributions can be used to produce more precise host-subtracted supernova spectra and better constrain their UV continuum slopes, enabling further analysis.

An approximate rate calculation of $z \sim 2$ SLSNe based on the SHIZUCA photometric candidate catalog is discussed in M19.  The accuracy of this rate is directly dependent on the efficiency of SHIZUCA in selecting high redshift transients and reducing photometric redshift confusion.  More spectra are needed to establish and improve this efficiency and separate high redshift supernovae from other high redshift transient phenomena such as AGN and TDEs.  As a first approximation, 3 of 5 SHIZUCA photometric candidates with spectroscopically confirmed redshifts (excluding the non-detected HSC17davs) are at high redshift, which gives a current redshift efficiency of 60\%.

SHIZUCA is capable of detecting supernovae to $z>6$, and with the right observing strategy and clear skies, spectroscopic follow-up of these events out to the edge of the epoch of reionization may be achievable.  And the potential of the James Webb Space Telescope will be greater still, capable of acquiring spectra of SLSNe as far as $z=20$.  At such high redshifts only deep, wide-field infrared surveys will be capable of producing targets.  Such surveys are already being considered in future facilities like the University of Tokyo Atacama Observatory (TAO) and the Kunlun Dark Universe Survey Telescope (KDUST).  By exploring the current limits of high redshift transient astronomy, SHIZUCA and similar programs are setting the stage for observing the explosions of the very first stars.

\acknowledgements
\begin{center}
ACKNOWLEDGEMENTS
\end{center}

This research is supported by the Grants-in-Aid for Scientific Research of the Japan Society for the Promotion of Science (TJM 16H07413, 17H02864) and by JSPS Open Partnership Bilateral Joint Research Project between Japan and Chile.

C.C. would like to express his appreciation to Uros Mestric for all his assistance with the observations and the data analysis and to Mat Smith for his contribution of the reduced spectrum of DES15E2mlf.

C.C. would also like to thank Chuck Steidel, Robin Humble and Sam Hinton for their helpfulness in discussions of spectroscopic cross-correlation.

J.C. acknowledges the Australian Research Council Future Fellowship grant FT130101219 and the Australian Research Council Centre of Excellence for Gravitational Wave Discovery (OzGrav), CE170100004

L.G. was supported in part by the US National Science Foundation under Grant AST-1311862.

Support for G.P. is provided by the Ministry of Economy, Development, and Tourism's Millennium Science Initiative through grant IC120009, awarded to The Millennium Institute of Astrophysics, MAS.
G.P. also acknowledges support by the Proyecto Regular FONDECYT 1140352.

This research is partly supported by Japan Science and Technology Agency CREST JPMHCR1414.

Parts of this research were conducted by the Australian Research Council Centre of Excellence for All-sky Astrophysics (CAASTRO), through project number CE110001020.

The Hyper Suprime-Cam (HSC) collaboration includes the astronomical communities of Japan and Taiwan, and Princeton University.  The HSC instrumentation and software were developed by the National Astronomical Observatory of Japan (NAOJ), the Kavli Institute for the Physics and Mathematics of the Universe (Kavli IPMU), the University of Tokyo, the High Energy Accelerator Research Organization (KEK), the Academia Sinica Institute for Astronomy and Astrophysics in Taiwan (ASIAA), and Princeton University.  Funding was contributed by the FIRST program from Japanese Cabinet Office, the Ministry of Education, Culture, Sports, Science and Technology (MEXT), the Japan Society for the Promotion of Science (JSPS),  Japan Science and Technology Agency  (JST),  the Toray Science  Foundation, NAOJ, Kavli IPMU, KEK, ASIAA,  and Princeton University.
 
This paper makes use of software developed for the Large Synoptic Survey Telescope. We thank the LSST Project for making their code available as free software at \url{http://dm.lsst.org}.
  
Based on data collected at the Subaru Telescope and retrieved from the HSC data archive system, which is operated by the Subaru Telescope and Astronomy Data Center at National Astronomical Observatory of Japan.

Based in part on data obtained at the Gemini Observatory via the time exchange program between Gemini and the Subaru Telescope processed using the Gemini IRAF package (program ID: S17A-056, GS-2017A-Q-13). The Gemini Observatory is operated by the Association of Universities for Research in Astronomy, Inc., under a cooperative agreement with the NSF on behalf of the Gemini partnership: the National Science Foundation (United States), the National Research Council (Canada), CONICYT (Chile), Ministerio de Ciencia, Tecnolog\'{i}a e Innovaci\'{o}n Productiva (Argentina), and Minist\'{e}rio da Ci\^{e}ncia, Tecnologia e Inova\c{c}\~{a}o (Brazil).

This research has made use of the NASA/ IPAC Infrared Science Archive, which is operated by the Jet Propulsion Laboratory, California Institute of Technology, under contract with the National Aeronautics and Space Administration.

Some of the data presented in this paper were obtained from the Mikulski Archive for Space Telescopes (MAST). STScI is operated by the Association of Universities for Research in Astronomy, Inc., under NASA contract NAS5-26555. Support for MAST for non-HST data is provided by the NASA Office of Space Science via grant NNX09AF08G and by other grants and contracts.

Support for {\it HST} program GO-13784 was provided by NASA through a grant from the Space Telescope Science Institute, which is operated by the Association of Universities for Research in Astronomy, Inc., under NASA contract NAS 5-26555.

The authors wish to recognize and acknowledge the very significant cultural role and reverence that the summit of Mauna Kea has always had within the indigenous Hawaiian community.  We are most fortunate to have the opportunity to conduct observations from this mountain.

\facilities{Keck/LRIS, Subaru/HSC, Gemini/GMOS-S} 
\software{\texttt{MIZUKI} \citep{tanaka15}, \texttt{MARZ} \citep{hinton16}}


\bibliographystyle{yahapj}

\bibliography{mybib}



\end{document}